\documentclass[11pt,aps,amssymb,a4paper,prd,preprintnumbers,nofootinbib]{revtex4-1}
\pdfoutput=1
\usepackage{fullpage}
\usepackage{float}
\usepackage{amsfonts}
\usepackage{amsmath}
\usepackage{slashed}
\usepackage{amssymb}
\usepackage{graphicx}
\usepackage{makeidx}
\usepackage{cancel}
\usepackage{eepic}
\usepackage{epsfig}
\usepackage{natbib}
\usepackage{latexsym}
\usepackage[dvipsnames]{xcolor}
\usepackage{float}
\usepackage{multirow}
\usepackage[export]{adjustbox}
\usepackage{xurl,hyperref}
\usepackage{enumitem}
\usepackage{float}
\newcommand{\beq}{\begin{eqnarray}}
\newcommand{\eeq}{\end{eqnarray}}
\def\lsim{\;\raise0.3ex\hbox{$<$\kern-0.75em\raise-1.1ex\hbox{$\sim$}}\;}
\def\gsim{\;\raise0.3ex\hbox{$>$\kern-0.75em\raise-1.1ex\hbox{$\sim$}}\;}
\def\beq{\begin{equation}}   \def\eeq{\end{equation}}
\def\ba{\begin{array}}       \def\ea{\end{array}}
\def\bea{\begin{eqnarray}}   \def\eea{\end{eqnarray}}

\def\k{\kappa}
\def\l{\lambda}

\begin{document}
\qquad\qquad\preprint{IP/BBSR/2023-08}

\title{ Testing $Z$ boson rare decays $Z\to H_1 \gamma, A_1 \gamma$ with $(g-2)_\mu$, $ M_W$, and $BR(h_{\rm SM}\to Z\gamma)$ in the NMSSM}
\author{Subhadip Bisal}
\email{subhadip.b@iopb.res.in}
\affiliation{Institute of Physics, Sachivalaya Marg, Bhubaneswar 751 005, India}
\affiliation{Homi Bhabha National Institute, Training School Complex, Anushakti Nagar, Mumbai 400 085, India}
\author{Debottam Das}
\email{debottam@iopb.res.in}
\affiliation{Institute of Physics, Sachivalaya Marg, Bhubaneswar 751 005, India}
\affiliation{Homi Bhabha National Institute, Training School Complex, Anushakti Nagar, Mumbai 400 085, India}

\date{\today}

\begin{abstract}
\noindent 
We study the rare decay process of $Z$ boson into photon, accompanied by a CP-even or CP-odd scalar. We present the analytical delineation of the processes through the model-independent parametrizations of the new physics couplings and, finally, consider the Next-to-Minimal Supersymmetric Standard Model to mark out the parameter space where the branching fraction can have the maximum value. 
As a part of the necessary phenomenological and experimental cross-checks, we aim to fit the anomalous magnetic moment of the muon and $W$ boson mass anomaly through the supersymmetric contributions. We also find that the decays $Z\to H_1 \gamma, A_1 \gamma$ can serve as an excellent complementary test to $BR(h_{\rm SM}\to Z\gamma)$. In fact, to facilitate future searches, we unveil a few benchmark points that additionally satisfy the deviation of $BR(h_{\rm SM}\to Z\gamma)$ from the SM value based on the recent measurements of ATLAS and CMS.
Future proposals such as ILC, CEPC, and FCC-ee are anticipated to operate for multiple years, focusing on center-of-mass energy near the $Z$ pole. Consequently, these projects will be capable of conducting experiments at the Giga-$Z$ ($10^{9}$ of $Z$ bosons) and Tera-$Z$ ($10^{12}$ of $Z$ bosons) phases, which may probe the aforesaid rare decay processes, thus the model as well. These unconventional yet complementary searches offer different routes to explore the supersymmetric models with extended Higgs sectors like NMSSM.
\end{abstract}

\maketitle
\section{Introduction}
After the discovery of the Higgs boson at LHC in 2012 \cite{ATLAS:2012yve, CMS:2012qbp}, many precise measurements of its quantum numbers and couplings showed remarkable consistency with that of the Standard Model (SM) predictions \cite{ATLAS:2019nkf}, which include the loop-induced three-point vertices, e.g., $h\gamma\gamma$ or $ h Z \gamma$. The latter may be of importance to test the imprints of new physics through the rare decays of the $Z$ boson in the extended Higgs scenarios where an additional CP-even ($H_1$) or CP-odd ($A_1$) Higgs scalar (would be collectively referred to as $\phi$), lighter than the $Z$ boson can be present. In this regime, the other exotic $Z$ boson decays are $Z \to \bar{f} f \phi$ ($f$ refers to an SM fermion), and $Z\to 3 A_1$. None of the decays is possible in the SM.
However, with a lighter Higgs, the rates can be significantly enhanced \cite{Cahn:1978nz, Kalinowski:1988sp,
  Weiler:1988xn, Djouadi:1996yq, Krawczyk:2000kf, Krawczyk:1998kk, Djouadi:1999gv, Cao:2010na, Chang:1996er,
  Keith:1997ci, Perez:2003ad, Larios:2001ma, Larios:2002ha}.
  For instance, the decays of $Z$ boson into light Higgs bosons were studied, through $Z \to \bar{f} f \phi $ ($f=b, \tau$) in \cite{Kalinowski:1988sp,Krawczyk:2000kf,Cao:2010na}. The decay $Z \to \phi \gamma $ was studied in the extensions of the SM including Minimal Supersymmetric Standard Model (MSSM) \cite{Krawczyk:2000kf,  Weiler:1988xn, Djouadi:1996yq, Krawczyk:1998kk, Cao:2010na}.  Similarly, the decay of $Z$ boson into light pseudoscalars was studied in a model-independent way \cite{Chang:1996er}, in 2HDM or supersymmetric (SUSY) models \cite{Keith:1997ci, Perez:2003ad, Larios:2001ma, Larios:2002ha, Djouadi:1999gv, Cao:2010na}.

  Here we investigate the decay process $Z\rightarrow \phi\gamma$ in the context of the Next-to-Minimal Supersymmetric Standard Model (NMSSM).
The new scalars $H_1$ and $A_1$ may become lighter than $M_Z$ 
  in some parts of the NMSSM. In fact,   
  there exist limits like PQ-symmetry-limit ($\kappa \to 0$) or
R-symmetry-limit ($A_\kappa, A_\lambda \to 0$), where a light CP-odd Higgs bosons
would follow naturally \cite{Ellwanger:2009dp}. 
A light scalar helps to
accomplish different goals : (i)
to
satisfy the dark matter abundance of the Universe through resonance
annihilations of light Singlino-like or Bino-like DM pairs, (ii) to
initiate a strong first-order electroweak phase transition,
which also yields the promising conditions
for electroweak baryogenesis
\cite{Profumo:2007wc,Carena:2011jy, Baum:2020vfl,Carena:2022yvx,Chatterjee:2022pxf,Borah:2023zsb}, (iii) to boost the BSM contributions to the
anomalous magnetic moment of the muon \cite{Chang:2000ii, Feng:2008cn, Heinemeyer:2004yq, Domingo:2008bb} or the $W$ boson mass
at one-loop \cite{Domingo:2022pde, Stal:2015zca, Cao:2008rc} or even at the
two-loop \cite{Stal:2015zca}. Thus, the quest of probing light scalars is of interest, especially since this
is one of the salient features of the NMSSM.

Apart from fermions and gauge bosons in the SM, the charged Higgs scalars and charginos in the NMSSM will contribute through the virtual corrections
in the \textbf{$BR(Z\rightarrow \phi\gamma)$}.
Among the SM particles, the loops mediated by $W$ boson and $t$ quark will have the largest contributions for a CP-even Higgs scalar. But for the CP-odd scalar, only $t$ quark loop dominates. For a singlet-like $H_1(A_1)$, Higgsino-like chargino loop through the Yukawa term $\lambda H_1 \tilde{H_u}^+ \tilde{H_d}^- (\lambda A_1 \tilde{H_u}^+ \tilde{H_d}^-)$, can contribute significantly if $ \lambda \sim \mathcal O(1)$. Besides the loop induced decays, $Z \to \bar{f} f \phi $ ($f=b, \tau$) or $Z\to A_1 A_1 A_1$ can be generated at the tree-level \cite{Cao:2010na} \footnote{We also note that the leptonic $Z$ decays in the NMSSM were considered earlier \cite{Domingo:2011uf}. Similarly, the Higgs boson decays into $\gamma$, and a $Z$ boson at the one-loop level in the NMSSM has been considered \cite{Belanger:2014roa}.}. 
The Higgs-mediated loops were also considered for the latter process. The partial decay widths of the processes strongly depend on the doublet component in $H_1$ or $A_1$. Following the LHC results, the doublet parts are already constrained considerably. Assuming it will be curbed further at the future runs of LHC, the rare decays become less attractive. On the contrary, the dominant NMSSM contributions for the one-loop amplitude $Z \to \phi \gamma$ do not receive any suppression due to the composition in $\phi$. Thus we find it appropriate to study the upper limit of the branching ratio (BR) of $Z \to \phi \gamma$ in the NMSSM parameter space. With a large branching fraction, we may even probe a dominantly singlet-like scalar directly at the LHC. Otherwise, the leading order (LO) production cross-section, mediated by gluon-fusion, goes with the square of the tiny doublet component, thus highly suppressed. Even the Next-to-Leading Order (NLO) corrections to the cross-section cannot provide much boost for the NMSSM \cite{Bisal:2022nbn}.

After having calculated the $Z\to\phi\gamma$ amplitude, we highlight the parametric dependence of its branching fraction in terms of the model parameters obeying the necessary theoretical and experimental bounds. Moreover, we delineate the regions accompanying $BR(Z\to\phi\gamma)$ that can simultaneously explain
(i) the anomaly in the magnetic moment of the muon ($\delta a_\mu$), (ii) the anomaly in the measurements of $W$ boson mass ($\delta M_W$), (iii) dark matter (DM) in the form of the lightest SUSY particle (LSP) or lightest neutralino\footnote{The DM constraint is partially respected in this study.}, and (iv) the recent measurements of the loop-induced decay of an SM-like Higgs scalar into $Z$ boson and $\gamma$.
The hadronic contributions are often referred to raise doubts about the beyond-SM (BSM) contributions to fit the first two precision measurements independently, i.e., $5.1\sigma$ deviation of the long-standing muon $g-2$ anomaly and the recent CDF measurement of the $W$ mass, which corresponds to a 6$\sigma$ deviation. Importantly, two anomalies pull the hadronic contributions in opposite directions \cite{Athron:2022qpo}. Thus, a BSM interpretation is worth needed for simultaneous explanations of both. In the NMSSM, the new world average of $M_W$ with the anomalous magnetic moment of the muon and the DM relic density constraints have been studied recently \cite{Domingo:2022pde, Tang:2022pxh}. In this analysis, we would go one step further by connecting the low energy precision observables with the rare decays of $Z$ boson.
This, in turn, will help to probe the NMSSM parameter space, compatible with
$\delta a_\mu$ and $\delta M_W$ at the future $Z$-factories, based on the proposed high-energy $e^+e^-$ accelerators CEPC \cite{CEPCStudyGroup:2018rmc, CEPCStudyGroup:2018ghi}, FCC-ee \cite{FCC:2018evy}, or ILC \cite{Bambade:2019fyw, Erler:2000jg, Baer:2013cma}, which will operate in Giga-$Z$ ($10^9$ of $Z$s) and Tera-$Z$  ($10^{12}$ of $Z$s) phases.
 Additionally, here we consider the lighter electroweakinos and sleptons to explain the above anomalies and to have the maximum branching ratios for $Z \to \phi \gamma$.
Consequently, a few benchmark points (BMPs) have been proposed for illustration.

The rest of the work is organized as follows. We detail the $Z\rightarrow \phi \gamma$ process in Sec.~\ref{sec:zphigamma_amp}. Then we outline the relevant aspects of the NMSSM in Sec.~\ref{sec:NMSSM}. 
Here we mainly consider the electroweak (EW) particles, e.g., charginos,  sleptons, and charged Higgs in the loops. Note that squarks or sleptons contribute only for $Z\rightarrow H_1 \gamma$ \cite{Djouadi:1996yq}. 
Sec.~\ref{sec:results} covers the choice of the parameters and constraints from theory and experiments relevant to the present purpose. We detail all the precision observables in this regard.
We present the relevance of all the considered loops numerically in Sec.~\ref{Sec:NR}.
Numerical results are also elaborated in this section. Finally, we conclude in Sec.~\ref{Sec:conclusion}.

\section{Feynman diagrams and the amplitude of $Z\rightarrow \phi\gamma$}
\label{sec:zphigamma_amp}

The amplitude for the decay $Z\rightarrow \phi \gamma$
can, in general, be expressed as 
\begin{align}
  \mathcal{M}(Z\rightarrow \phi\gamma) &= \mathcal{M}(Z_\mu(p_1),A_\nu(p_2),\phi(p_3))
  \epsilon_1^{\mu}(p_1)\epsilon_2^{\nu *}(p_2)\nonumber\\
&= \mathcal{M}_{\mu\nu}\epsilon_1^\mu \epsilon_2^{\nu *}~,
\label{amplitudeeqn}
\end{align} 
where $\epsilon_1^{\mu}$ and $\epsilon_2^{\nu *}$ are the polarization vectors for the $Z$ boson and the photon $\gamma$, respectively. Here $p_1$, $p_2$, and $p_3$ are the external momenta satisfying the condition $p_1=p_2+p_3$ and the on-shell conditions $p_1^2=M_Z^2$, $p_2^2=0$, and $p_3^2= M_{\phi}^2$. The amplitude
$\mathcal{M}_{\mu\nu}$ is generally written in the following form :
\begin{align}
  \mathcal{M}_{\mu\nu} = \mathcal{F}_{00}\, g_{\mu\nu} + \sum_{i,j=1}^{2} \mathcal{F}_{ij}p_{i\mu}p_{j\nu} +
  i\, \tilde{\mathcal{F}}\, \epsilon_{\mu\nu\rho\sigma}p_1^\rho p_2^\sigma~,
\end{align}
where $\mathcal{F}_{00}$, $\mathcal{F}_{ij}$ $(i,j=1,2)$ and $\tilde{\mathcal{F}}$ are the scalar form factors
which can be computed from the loop diagrams and $\epsilon_{\mu\nu\rho\sigma}$ is the totally antisymmetric
tensor with $\epsilon_{0123}=-1$ and $\epsilon^{0123}=+1$. 
Since the on-shell photon satisfies the condition $\epsilon_2^{\nu *}p_{2\nu}=0$, one finds that $\mathcal{F}_{12}$
and $\mathcal{F}_{22}$ do not contribute to the amplitude in Eq.~\eqref{amplitudeeqn}. Also, the Ward identity
implies that $p_2^{\nu}\mathcal{M}_{\mu\nu}=0$, which results $\mathcal{F}_{11}=0$ and 
\begin{align}
\mathcal{F}_{00} = -(p_1.p_2)\mathcal{F}_{21}= \frac{M_Z^2-M_{\phi}^2}{2}\mathcal{F}_{21}~.
\end{align}
\noindent
Therefore, the amplitude can be cast as
\begin{align}
  \mathcal{M}_{\mu\nu} = \mathcal{F}\big[-(p_1.p_2)g_{\mu\nu}+ p_{2\mu}p_{1\nu}\big]
  + i\,\mathcal{\tilde{F}}\,\epsilon_{\mu\nu\rho\sigma}p_1^{\rho}p_2^{\sigma}~,
  \label{Eq:zha_tensorial}
\end{align}
where we have written $\mathcal{F}_{21}=\mathcal{F}$.
Thus it suffices to collect the coefficient of $p_{2\mu}p_{1\nu}$ to find the form factor $\mathcal{F}$. Note that, $\mathcal{F}$ is relevant for $Z\to H_1\gamma$ while 
$\mathcal{\tilde{F}}$ 
contributes to $Z\to A_1\gamma$ process
only.

\begin{figure}[H]
	\centering
	\includegraphics[width=0.7\linewidth]{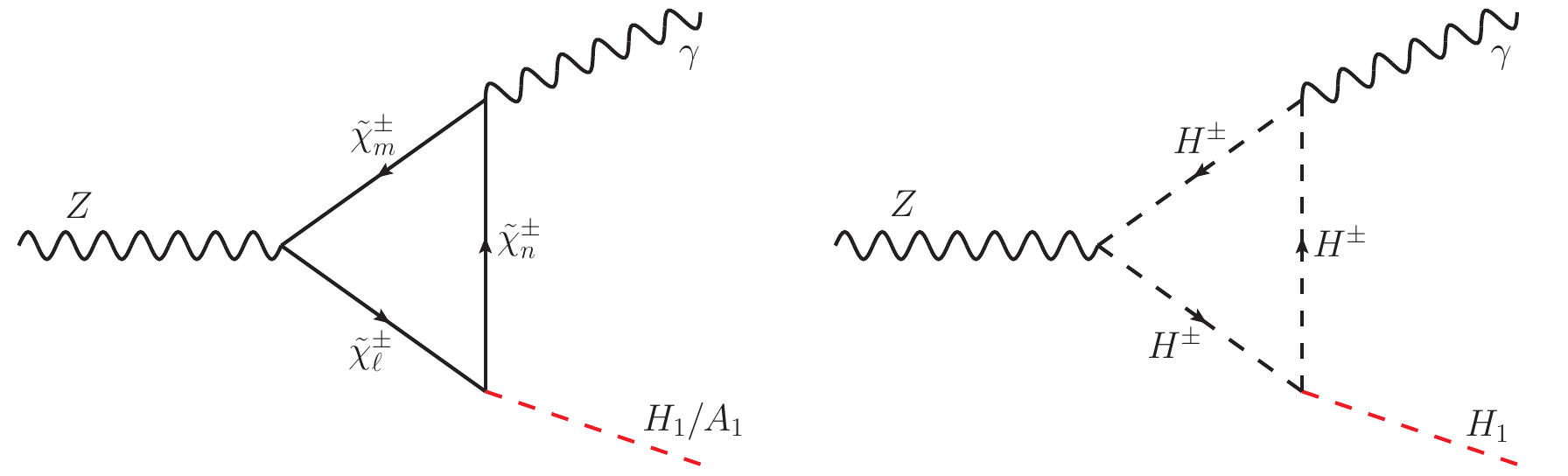}\\
	\quad(a)\qquad\qquad\qquad\qquad\qquad\qquad\qquad(b)
	\vskip 0.8cm
	\includegraphics[width=0.7\linewidth]{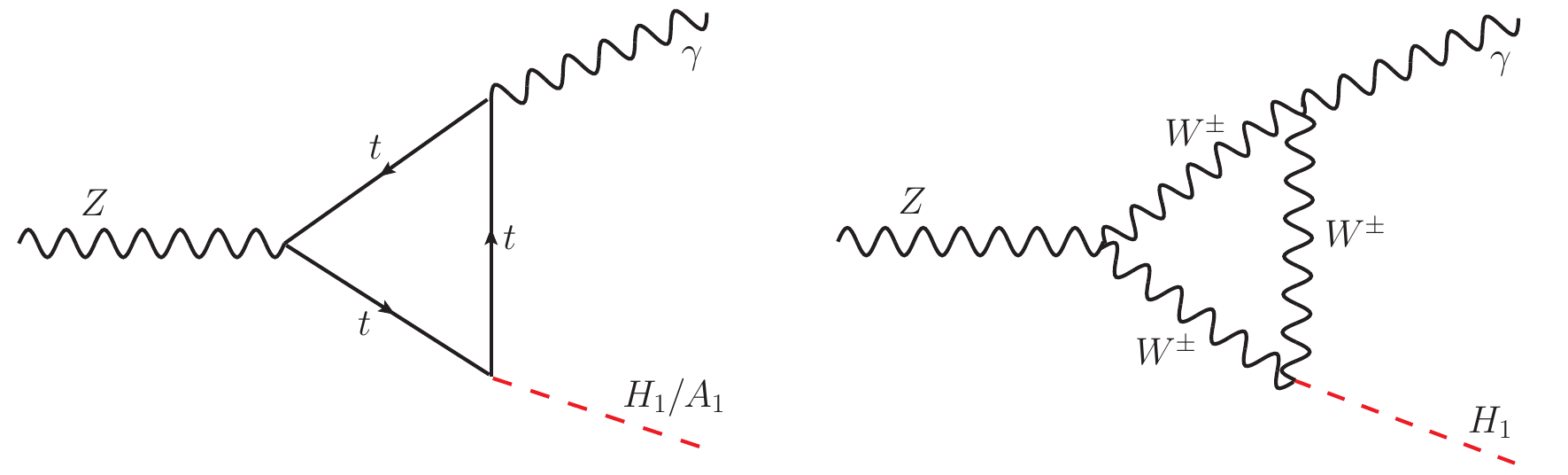}\\
	\quad(c)\qquad\qquad\qquad\qquad\qquad\qquad\qquad(d)
	\vskip 0.8cm
	\includegraphics[width=0.4\linewidth]{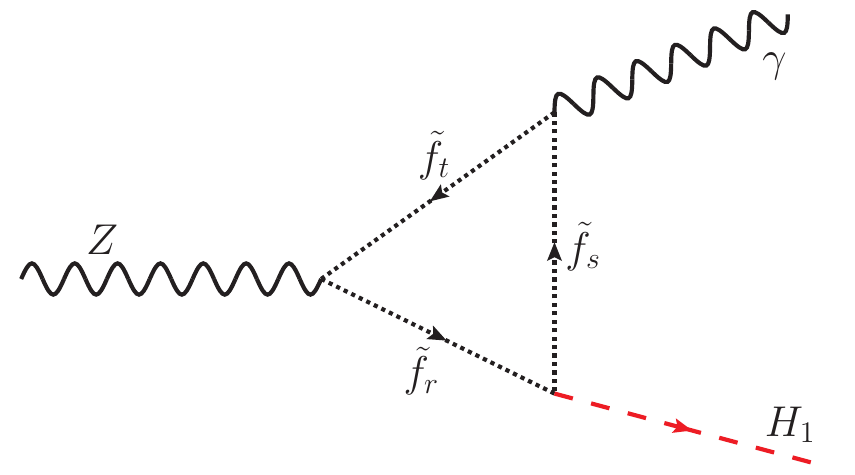}
	\includegraphics[width=0.4\linewidth]{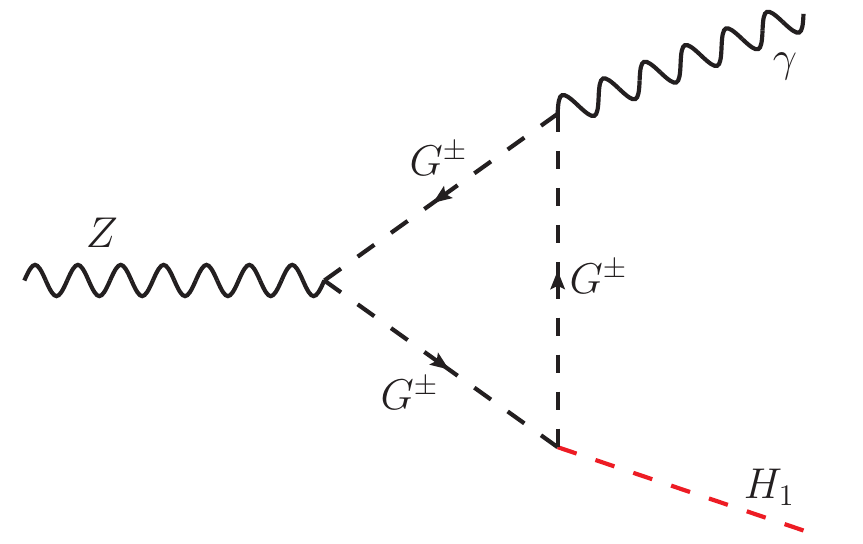}\\
	\quad(e)\qquad\qquad\qquad\qquad\qquad\qquad\qquad(f)
	\caption{One-loop Feynman diagrams contributing to the process $Z\rightarrow \phi\gamma$ ($\phi=H_1,A_1$) in the NMSSM.}
	\label{fig:zagamma3} 
\end{figure}

For a specific realization, we consider the scale-invariant NMSSM. Considering Feynman's gauge for this
analysis, the relevant Feynman diagrams have been shown in Fig.~\ref{fig:zagamma3}. 
However, the fermion and $W$ boson loops are separately gauge invariant if all the external particles are on–shell \cite{Djouadi:1996yq}.
We use Package-X \cite{Patel:2016fam, Patel:2015tea} to express the Feynman amplitudes in terms of the Passarino-Veltman integrals.  
In Eq.~\eqref{formfactor:charginoloop}-\eqref{formfactor:goldstone}, the scalar parts of 
all the BSM couplings to the SM and BSM particles are noted as free parameters so as to have the amplitudes in a model-independent general form. One may readily read them from the NMSSM vertices presented in the Appendix (Sec.~\ref{sec:appendixA}).

{\bf Fig.~\ref{fig:zagamma3}a, \ref{fig:zagamma3}c:} The amplitudes, mediated via the spin $1/2$ particles, are calculated and expressed as
\begin{align}
\mathcal{F}^{(a)} = &\frac{i}{16\pi^2}eg_{Z\tilde{\chi}_\ell^\pm\tilde{\chi}_m^\mp}g_{H_1\tilde{\chi}_\ell^\pm\tilde{\chi}_n^\mp}\delta_{mn}K^{(+)}_{LR} \Big[4\Bigl\{2\big(M_\ell+M_n\big)\big(\mathbf{C}_{12}+\mathbf{C}_{22}\big)+\big(M_\ell-M_m\big)\mathbf{C}_1\nonumber\\
&+\big(3M_\ell+M_n\big)\mathbf{C}_2+M_\ell \mathbf{C}_0\Bigr\}\Big]~,
\label{formfactor:charginoloop}\\
\mathcal{F}^{(c)}
= &\frac{ig_2e}{16\pi^2c_W } L_{LR}^{(+)}g_{H_1tt}M_t \Big[8\big(\mathbf{C}_{12}+\mathbf{C}_{22}\big)+8\mathbf{C}_2+2\mathbf{C}_0\Big]~, 
\label{fig:1ac}
\end{align}

and for the pseudoscalar, the form factors are given by 
\begin{align}
    \tilde{\mathcal{F}}^{(a)} = &-\frac{1}{16\pi^2} e g_{Z\tilde{\chi}_\ell^\pm\tilde{\chi}_m^\mp}g_{A_1\tilde{\chi}_\ell^\pm\tilde{\chi}_n^\mp}\delta_{mn}K^{(+)}_{LR} \Big[4\Bigl\{M_{\ell}\mathbf{C}_{0} + \big(M_{\ell}-M_m\big)\mathbf{C}_1 \nonumber+ \big(M_{\ell}-M_n\big)\mathbf{C}_2\Bigr\}\Big]~,\\
    \tilde{\mathcal{F}}^{(c)} = &-\frac{g_2e}{16\pi^2c_W } L_{LR}^{(+)}g_{A_1tt}M_t \big(2\mathbf{C}_0\big)
\end{align}
In the above, $K_{LR}^{(\pm)} = \mathcal{O}^{L}_{m\ell}\pm \mathcal{O}^{R}_{m\ell}$~, with $\mathcal{O}^{L,R}_{m\ell}$ are defined in the Appendix. 
Similarly, $L_{LR}^{\pm}=C_L\pm C_R$ with $C_L=2-\dfrac{8}{3}s_W^2$ and $C_R=-\dfrac{8}{3}s_W^2$.

{\bf Fig.~\ref{fig:zagamma3}b, \ref{fig:zagamma3}d-\ref{fig:zagamma3}f:} The amplitudes with integral spin particles within the loops can be cast as 
\begin{align}
\mathcal{F}^{(b)} = & -\frac{i}{16\pi^2}e g_{ZH^\pm H^\mp}\lambda_{H_1 H^\pm H^\mp}\Big[8\big( \mathbf{C}_{12}+\mathbf{C}_{22}+\mathbf{C}_2\big)\Big]~, \\
\mathcal{F}^{(d)} = & -\frac{ieg_2 c_W}{16\pi^2} g_{H_1 W^\pm W^\mp}\Big[2\Bigl\{(6-4d)\big(\mathbf{C}_{12}+\mathbf{C}_{22}+\mathbf{C}_2\big) -\mathbf{C}_1-5\mathbf{C}_0\Bigr\}\Big]~,\\
\mathcal{F}^{(e)} =&-\frac{i}{16\pi^2}eg_{H_1\tilde{f}_r\tilde{f}_s}g_{Z\tilde{f}_r\tilde{f}_t}\delta_{st}\Big[8\big(\mathbf{C}_{12}+\mathbf{C}_{22}+\mathbf{C}_2\big)\Big]~,\\
\mathcal{F}^{(f)} =& -\frac{i}{16\pi^2}eg_{ZG^\pm G^\pm}g_{H_1G^\pm G^\pm}\Big[8\big(\mathbf{C}_{12}+\mathbf{C}_{22}+\mathbf{C}_2\big)\Big]~.
\label{formfactor:goldstone}
\end{align}
The PV-functions are $\mathbf{C}_{ij}= \mathbf{C}_{ij}(M_Z^2,0,M_{\phi}^2; M_a,M_b,M_c)$ and $\mathbf{C}_{i}= \mathbf{C}_{i}(M_Z^2,0,M_{\phi}^2; M_a,M_b,M_c)$, where $a,b,c$ refer to particles within the loop. 
Similarly, $g_{ZW^\pm W^\mp}$ $=g_2 c_W$ etc. and $d$ is the dimensionality of space-time. Following the same way, one can easily obtain the amplitude for the decay process $H_2\to Z\gamma$, where $H_2$ is the SM-like Higgs in the NMSSM with $M_{H_2}\approx 125$ GeV.

The numerical values of the form factors are verified to the respective amplitudes, calculated using Feynman parameters in the Appendix (Sec.~\ref{Eq:loop_feynman}). Finally, the partial decay width for $Z\rightarrow \phi \gamma$ can be calculated as follows,
\begin{align}
\Gamma\big(Z\rightarrow \phi \gamma\big) = \frac{M_Z^3}{32\pi}\Bigg(1-\frac{M_{\phi}^2}{M_Z^2}\Bigg)^3\big(\lvert \mathcal{F}\rvert^2 + \lvert \mathcal{\tilde{F}}\rvert^2\big)~.
\label{Eq:width_form}
\end{align}
This rate is normalized to the total decay width $\Gamma_Z \simeq 2.5$ GeV.

\section{Contributions of $\mathcal{{F}}$ and $\mathcal{\tilde{F}}$ in the NMSSM}
\label{sec:NMSSM}

In the popular version of the NMSSM, the Higgs mass term $\mu \widehat H_u \widehat H_d$ in the MSSM
superpotential $W_{\text{MSSM}}$ has been replaced by the coupling $\lambda$ of $\widehat S$ to $\widehat H_u$ and $\widehat H_d$, and a self-coupling $\kappa \widehat S^3$ \cite{Ellwanger:2009dp, Maniatis:2009re}. In this realization the superpotential $W_{\text{NMSSM}}$ becomes scale-invariant, and given by:
\beq\label{eq:1}
W_{\text{NMSSM}} = \lambda \widehat S \widehat H_u\cdot \widehat H_d +
\frac{\kappa}{3}  \widehat S^3 +\dots\; ,
\eeq
where the dots denote the Yukawa couplings of the MSSM. The NMSSM-specific soft SUSY breaking terms consist of a mass term for the scalar components of $\widehat S$, and the trilinear interactions associated with the terms in $W_{\text{NMSSM}}$:
\beq\label{eq:2}
 -{\cal L}_{\text{NMSSM}}^{\rm Soft} = m_{S}^2 |S|^2 +\Bigl( \lambda
A_\lambda\, H_u \cdot H_d \,S +  \frac{1}{3} \kappa  A_\kappa\,  S^3
\Bigl)\ +\ \mathrm{h.c.}\;.
\eeq
Subsequently, the weak eigenstates $H_{d,u}$ and $S$ can be expanded around the vacuum expectation values (vevs) as defining $v^2 = 2M_Z^2/(g_1^2 + g_2^2) \sim (174 \rm GeV)^2)$, $\left<H_u\right>=v_u,\,\left<H_d\right>=v_d,\, \,{\rm and}\,\left<S\right>=s\;$. In terms of $s$, the first term in $W_{\text{NMSSM}}$ generates an effective $\mu$-term with
\beq\label{eq:4}
\mu_\mathrm{eff}=\lambda s\; .
\eeq
The general expressions for these mass matrices, including the dominant radiative corrections, are given in Ref. \cite{Ellwanger:2009dp}. Using the minimization equations of the potential to eliminate the soft SUSY breaking Higgs mass terms, the Higgs sector of the NMSSM can be characterized by six parameters at the tree-level.
\beq\label{eq:5}
\lambda\; ,\quad \kappa\; ,\quad A_\lambda\; ,\quad A_\kappa\; ,\quad
\mu_\mathrm{eff}\; , \quad \tan\beta \equiv \frac{v_u}{v_d}\; .
\eeq
The neutral CP-even Higgs sector contains three states $H_i\,(i=1,2,3)$, which are the mixtures of the CP-even components of the superfields $\widehat H_u$, $\widehat H_d$, and $\widehat S$. Their masses are described by a $3 \times 3$ mass matrix ${\cal M}^2_{H\,ij}$, where the dominant contribution to the singlet-like component ${\cal M}^2_{H\,33}$ reads \cite{Maniatis:2009re,Ellwanger:2009dp} ($s >> v_u, v_d$)
\beq\label{eq:cpeven}
{\cal M}^2_{H\,33} \sim \kappa s\left(A_\kappa + 4 \kappa s\right)\; .
\eeq
The neutral CP-odd Higgs sector contains two physical states $A_i\,(i=1,2)$, whose masses are described by a $2 \times 2$ mass matrix ${\cal M}^2_{A\,ij}$ where ${\cal M}^2_{A\,11}$ corresponds to the MSSM-like CP-odd Higgs mass squared. The dominant contribution to the singlet-like component ${\cal M}^2_{A\,22}$ is given by \cite{Maniatis:2009re,Ellwanger:2009dp}, 
\beq\label{eq:cpodd}
{\cal M}^2_{A\,22} \sim -3\kappa s A_\kappa ~. 
\eeq
Following 
Eq.\ref{eq:cpeven} or Eq.\ref{eq:cpodd}, we may say
that $\kappa \to 0$ can indeed
be helpful to produce a light $H_1$ or $A_1$.
In the neutralino sector, we have five neutral states ${\tilde\chi}^0_i\,(i=1,2,\dots,5)$, which are the mixtures of Bino ($\tilde{B}$), neutral Wino ($\tilde{W}^3$), neutral Higgsinos ($\tilde{H}_d^0$ and $\tilde{H}_u^0$) from the superfields $\widehat H_d^0$ and $\widehat H_u^0$, and Singlino ($\tilde{S}$) from the superfield $\widehat S$. Their masses are described by a symmetric $5 \times 5$ mass matrix ${\cal M}_{\tilde{\chi}^0_{ij}}$ given by,
\beq\label{eq:neumass}
{\cal M}_{\tilde{\chi}^0_{ij}} =
\left( \ba{ccccc}
M_1 & 0 & -\frac{g_1 v_d}{\sqrt{2}} & \frac{g_1 v_u}{\sqrt{2}} & 0 \\
& M_2 & \frac{g_2 v_d}{\sqrt{2}} & -\frac{g_2 v_u}{\sqrt{2}} & 0 \\
& & 0 & -\mu_\mathrm{eff} & -\l v_u \\
& & & 0 & -\l v_d \\
& & & & 2 \k s
\ea \right)~,
\eeq
where $M_1$ and $M_2$ are the soft SUSY breaking mass terms for the Bino and Wino, respectively. In an R-parity conserving NMSSM, the lightest eigenstate $\tilde{\chi}^0_1$ is a valid DM candidate for most parts of the parameter space. With an orthogonal real matrix $N_{ij}$ that diagonalizes ${\cal M}_{\tilde\chi^0_{ij}}$, the composition of $\tilde{\chi}^0_1$ reads as
\beq\label{eq:neudecom}
\tilde{\chi}^0_1=N_{11} \tilde{B} + N_{12} \tilde{W}^3 + N_{13} \tilde{H}^0_d
 + N_{14} \tilde{H}^0_u + N_{15} \tilde{S}\; .
 \eeq
 Our requirement to boost the process $Z \to \phi \gamma$ via NMSSM-specific Higgsino-like chargino leads to a scenario where $\tilde{\chi}_1^\pm \simeq \tilde{\chi}_2^0$ is dominantly Higgsino-like.
 We will see that other EW observables
like $a_\mu$ or $W$ boson mass anomaly also favor a light chargino. The latter observables may receive boosts from other SUSY particles, e.g., through
lighter $\tilde \mu$, or lighter neutralinos. However,
$\tilde \mu$ does not have
any significance to $BR(Z \to \phi \gamma)$ while $\tilde \chi_i^0$ is not
involved in the $Z$ boson rare decay. 
But connecting all of them on the same thread as
done in this work, we will also be able to probe the parameter space with lighter EW spectra in general.
 Additionally, compliance with the observed relic abundance data and spin-independent (SI) DM-nucleon direct detection (DD) cross-section critically restrict $\tilde{\chi}_1^0$ to become Higgsino-like in the NMSSM parameter space (see Sec.~\ref{sec:results}). Here a dominantly Singlino-like DM is favoured.

We now qualitatively discuss the relative contributions of different loops of Fig.~\ref{fig:zagamma3}. The numerical verification will follow in Table \ref{tab:my_label_1} for a few representative
values of the input parameters. Among the different loops, the $W$ boson loop still carries the dominant
contribution for a singlet-like $\phi$ ($W$ boson loop would be absent for $A_1$) followed by the fermion loops through lighter charginos. The latter, induced by $\lambda \widehat S \widehat H_u\cdot \widehat H_d$ in the superpotential \eqref{eq:1}, can be seen to be suppressed up to one order of magnitude. Third-generation $t$ quark loop comes next in the relative contributions though it interferes destructively with the $W$ boson loop.
The contributions from the light fermions are suppressed via their masses. For sleptons, in particular, even for small $\tilde \mu$ masses ($M_{\tilde{\mu}}\leq 200$~GeV), the respective form factors  $\mathcal{F}^{(e)}_{\tilde{\mu}}$, $\mathcal{F}^{(e)}_{\tilde{\tau}}$ (superscripts refer to subfigures in Fig.~\ref{fig:zagamma3}) are much suppressed for any consideration. The same is true for the other loops. We note in passing that since $Z \to A_1\gamma$ amplitude is fully controlled by the chargino loops, we can measure the coupling $\lambda$ with its observance.

\section{Choice of parameters and constraints}
\label{sec:results}
\noindent
We use $ \mathtt{NMSSMTools-6.0.0}$ \cite{Ellwanger:2004xm, Ellwanger:2005dv, Das:2011dg} to compute the masses and couplings for SUSY particles, Higgs bosons, and the branching ratios (BRs) of the $Z$ boson decays.
For the latter observables, we modify the $\mathtt{NMSSMTools}$ suitably. In the present analysis, we set squarks and gluino masses at $\geq$ 2.5~TeV and $\geq$~3~TeV to cope with the LHC constraints \cite{ATLAS:2020syg, CMS:2019zmd}. We consider small masses $\sim$ [100-400]~GeV for the sleptons.  
This is because the lighter smuons are needed to satisfy the anomalous magnetic moment of the muon or $(g-2)_\mu$. In the case of $Z\to A_1\gamma$, the squarks and sleptons contributions will not be there due to CP-invariance \cite{Djouadi:1996yq}. Similarly, a lighter stau may help to reduce the branching fraction of $\tilde \chi^\pm_1$ in the classical electroweakino search channel. It may also produce friendly contributions to $\delta M_W$ \cite{Domingo:2022pde}. Here, we summarize different avenues of precision and collider phenomenology, which can be marked along with the $BR(Z\rightarrow \phi \gamma)$ in different parts of the NMSSM parameter space.

{\textbf{Lightest Neutralino as DM}} :
Since $\tilde{H}_u^+$ and $\tilde{H}_d^-$ of $\tilde{\chi}_i^\pm$ go with the Yukawa
coupling $\lambda H_1 \tilde{H}_u^+ \tilde{H}_d^-$ and $\lambda A_1 \tilde{H}_u^+ \tilde{H}_d^-$ of the singlet-like Higgs states, we require a large $\lambda$ and light Higgsino-like chargino states for favorable results.  
However, with a small mass, Higgsino fails to become a good DM candidate as larger Higgsino components of $\tilde{\chi}^0_1$ imply a too-small relic density or an underabundance of DM.
The acceptable value of the relic abundance data reads as \cite{WMAP:2012nax, Planck:2018vyg}%
 \begin{equation}
   \Omega_{\rm DM}h^2=0.1198\pm 0.0012.
   \label{eq:relic}
 \end{equation}
 Since Bino or Singlino can lead to an overabundance of DM, a reasonable choice is to consider either $\tilde B-\tilde H$ or $\tilde S-\tilde H$ type DM.
  But in the former case, DD experiments \cite{XENON:2018voc, LUX:2017ree, PandaX-II:2017hlx, LUX-ZEPLIN:2022xrq}, in particular, the recent XENONnT \cite{XENON:2020kmp} and LZ \cite{LUX-ZEPLIN:2022xrq} can lead to stringent constraints on the $\tilde{\chi}^0_1$-nucleon parameter space, which is particularly severe for the lighter Higgs spectra. A dominantly Bino-like LSP can satisfy the DM constraints but at the cost of larger values of the Higgsino mass parameter, which is not favoured in the present analysis. Note that the available parameter space would be further squeezed if one counts on the one-loop
  	corrections to three-point DM-DM-Higgs vertices involving EW SUSY particles~\cite{Bisal:2023fgb, Bisal:2023iip}. However, the non-Standard-Model-like Yukawa coupling \cite{Das:2020ozo} for an SM-like Higgs scalar with the first two generations of fermions may relax this tension. 

On the contrary, for the $\tilde S-\tilde H$ DM with a dominantly Singlino component, one can match the observed relic abundance and the DD bounds from ${\tilde\chi}^0_1$-nucleon cross-sections. This region usually 
refers to a lower value of $\lambda$ and $\kappa$. Without any surprise, the chargino contributions to $Z\rightarrow \phi \gamma$ will be subdued, which will affect the $BR(Z \rightarrow A_1 \gamma)$ more compared to the $BR(Z \rightarrow H_1 \gamma)$. For the latter, $W$ boson has significant parts in the partial amplitude, independent of the choice of $\lambda$ and $\kappa$.

{\textbf{A 125 GeV SM-like Higgs boson : }}
In the Higgs sector, the physical Higgs spectra yield three Higgs states: mostly ``singlet-like" ($H_1$), SM-like ($H_2$), and mostly ``MSSM-like" ($H_3$). The lightest CP-even Higgs state is $H_1$, while the same in the CP-odd sector is $A_1$. A large value for $\lambda$, desired to maximize the chargino contributions in $Z \to \phi \gamma$ amplitude, also helps to accommodate an SM-like Higgs boson $H_2$ with a mass of 125 GeV more easily. 
We also keep an eye on the SM-like Higgs boson couplings using the combined limits of ATLAS and CMS \cite{ATLAS:2016neq, CMS:2018uag, ATLAS:2020qdt, ATLAS-CONF-2021-053} in the conventional $\kappa$-frameworks. Specifically, with $\kappa_{H_iVV}$ (ratio of Higgs couplings to the vector bosons relative to the corresponding couplings of the SM-like Higgs boson) holds $ \sum_{i=1}^{3} \kappa^2_{H_iVV}=1$, there can only be a small allowance for a singlet-like or any of the BSM Higgs couplings to the vector bosons. Indeed considering $\kappa_{H_{125}VV} \sim 0.96$ at central value, $\kappa_{H_{\rm BSM}VV}$ can accommodate an allowance of $\sim 0.3$ (where $H_{\rm BSM}$ refers to any BSM Higgs scalars $H_1$ or $H_3$) \cite{ATLAS-CONF-2021-053}. We may primarily consider $M_{H_1}\,(M_{A_1})\gtrsim$ 60 GeV for the lightest scalar (pseudoscalar) so that the decay of an SM-like Higgs boson into a pair of lightest scalars (pseudoscalars) is kinematically forbidden. Otherwise, their couplings to the SM-like Higgs boson must be small enough, which can be possible when $\lambda$ and $\kappa$ assume smaller values. A 125 GeV SM-like Higgs boson can still be accommodated for the latter choice. 
Apart from the radiative corrections induced by the large values of $M_{\tilde{t}_1}$, the mixing between $H_1$ and $H_2$ or, more precisely, among the weak eigenstates leads to an increase in $H_2$ mass. A mass shift up to $\sim 8$~GeV can be achieved for larger $\tan\beta$ and smaller $\lambda$ values.

We always check the absence of unphysical global minima of the
Higgs potential as done in $\mathtt{NMSSMTools}$. We cross-checked
the validity of our BMPs with
$\mathtt{HiggsBounds-5.10.2}$ \cite{Bechtle:2020pkv, Bechtle:2008jh}.
Similarly, the compatibility of $H_2$ with the LHC
rate measurements for
the SM-like Higgs boson with a mass of 125 GeV has also been checked.
This is done using HiggsSignals-2.6.2 \cite{Bechtle:2013xfa,Bechtle:2020uwn}.
The signal strength predictions
for $H_2$ for the BMPs in TABLE \ref{tab:my_label} are within
the $\sim 1\sigma$ of each measurement. For estimation,
we have defined
$\chi^2_{red} = \frac{\chi^2}{n_{obs}}$
\cite{Biekotter:2019kde,Heinemeyer:2021msz}, where $\chi^2$
for a BMP obtained from HiggsSignals is
divided by the number of experimental
observations considered ($n_{obs}=111$). For our BMPs, we have obtained
$\chi^2_{red} \simeq 1$.

\textbf{$\mathbf{H_2}\to \mathbf{Z\gamma}$ in the NMSSM}: 
The inverse decay, but for the SM-like Higgs scalar, $H_2\to Z\gamma$ has a small branching fraction of about $1.5\times 10^{-3}$ in the SM \cite{Djouadi:1997yw}. Since this decay appears through loop diagrams, it opens the door to several BSM scenarios that may modify this branching fraction differently from the SM value.\\
The branching ratio for the decay $H_2\to Z\gamma$ is
recently measured, with a significance of $3.4\sigma$ \cite{ATLAS-CONF-2023-025}. A combination of ATLAS \cite{ATLAS:2020qcv} and CMS \cite{cmscollaboration2022search} search reports that the branching fraction for this decay channel is $(3.4\pm 1.1)\times 10^{-3}$ which is ($2.2 \pm 0.7$) times the SM prediction. Their analyses are based on Run-2 data sets collected by ATLAS and CMS experiments, corresponding to the $\int \mathcal{L}$ = 139 ${\rm fb}^{-1}$ and 138 ${\rm fb}^{-1}$ at $\sqrt{s}=13$ TeV. At present, it is quite early
 to state anything conclusively about its variation compared to the SM.
 However, if it gets settled in the future,
 then $BR({H_2}\to {Z\gamma})$ may become
 an interesting probe and complementary to $BR(Z\to H_1,A_1 \gamma)$
 to test some parts of the NMSSM parameter space.

\textbf{Precision observables in the NMSSM}: We impose the constraints from B-physics and recent results on the anomalous magnetic moment of the muon \cite{Muong-2:2021ojo, Muong-2:2021vma}, $\delta M_W$ \cite{Domingo:2011uf, Domingo:2022pde} as done in $\mathtt{NMSSMTools}$ \cite{Domingo:2007dx, Domingo:2008bb}. However, we find it essential to discuss some of them in some detail.

\textbf{Anomalous magnetic moment ($a_{\mu}$)}: The recent $a_\mu$ measurement by FNAL \cite{Muong-2:2021vma, Muong-2:2021ojo} has confirmed the earlier result by the E821 experiment at Brookhaven, yielding the experimental average $a_\mu^{\rm EXP}= (116592061\pm 41)\times10^{-11}$ which leads to a 4.2$\sigma$ discrepancy \cite{Muong-2:2021ojo} compared to the SM value $a_\mu^{\rm SM} = (116591810 \pm 43)\times 10^{-11}$ \cite{Aoyama:2020ynm}.
\begin{align}
\delta a_\mu = a_\mu^{\rm EXP}-a_\mu^{\rm SM} = (251\pm 59)\times10^{-11}.
\end{align}
Very recently, the E989 experiment at Fermilab released an update regarding the measurement of $a_\mu$ from Run-2 and Run-3. 
The new combined value yields a deviation of \footnote{The value of $(g-2)_\mu$ from Run-2 and Run-3 is $a_\mu^{\rm Run-2,3} = (116592055 \pm 24)\times 10^{-11}$. Therefore, the new experimental average becomes $a_\mu^{\rm Exp(New)} = (116592059 \pm 22)\times 10^{-11}$ \cite{Muong-2:2023cdq}.}
\begin{align}
   \delta a_\mu^{\rm New} = (249 \pm 48)\times 10^{-11} 
\end{align}
which leads to a $5.1\sigma$ discrepancy. 

It may be noted here that the lowest-order hadronic vacuum polarization contributions in $a_\mu^{\rm HVP}$ (hadronic loop contributions) is regarded as the main source of the
	present SM uncertainty of $a_\mu^{\rm SM}$.
	The HVP contribution is estimated in Ref.~\cite{Aoyama:2020ynm} as $a_\mu^{\rm HVP} \big\lvert_{e^+e^-}=(6931\pm40)\times 10^{-11}$ using $e^{+}e^{-}\to$ hadrons data. This differs significantly from the lattice-QCD (average) results available at that time, $a_{\mu}^{\rm HVP}\big\lvert_{\rm lattice}=(7116\pm184)\times 10^{-11}$, but is consistent with the uncertainties. BMWc later published a lattice-QCD result of $a_{\mu}^{\rm HVP}\big\lvert_{\rm BMWc}= (7075\pm55)\times 10^{-11}$~\cite{Borsanyi:2020mff}, reducing the discrepancy with $a_\mu^{\rm exp}$ to 1.6$\sigma$, but showing a 2.1$\sigma$ tension with the $e^+e^-$ determination. Other lattice groups~\cite{Ce:2022kxy, ExtendedTwistedMass:2022jpw, FermilabLatticeHPQCD:2023jof} somewhat agree with the BMW result. Overall, a significant disagreement among these observations concerning the HVP contributions may be easily noted. Furthermore, the difference between the two approaches used to determine the HVP contribution implies unresolved issues within each approach.
	
	The use of Euclidean time windows to address discrepancies between lattice-QCD and $e^+e^-\to$ hadrons cross-section data for the HVP contribution to $(g-2)_\mu$ is discussed in~\cite{Colangelo:2022vok}. Ref.~\cite{Wittig:2023pcl} also highlighted the tension between the lattice QCD approach and the traditional data-driven approach, while for the latter, the recent CMD-3 result was not used. Recently, $(g-2)_\mu$ has been calculated using the data-driven approach, measuring the $e^{+}e^{-}\to \pi^{+}\pi^{-}$ cross-section below 1 GeV with the CMD-3 detector, resulting in $\delta a_{\mu}= (49\pm 55)\times 10^{-11}$~\cite{CMD-3:2023rfe}. However, the recent data seem to be incompatible with previous determinations in~\cite{CMD-3:2023alj, CMD-2:2003gqi, KLOE-2:2017fda, BESIII:2015equ, BaBar:2012bdw}.
	Overall, we see significant differences between the HVP results among these different studies, suggesting further studies to resolve the discrepancy~\cite{Colangelo:2022jxc}.

Interestingly, $\delta a_\mu$ can be attributed to the SUSY contributions, which in turn helps us to constrain the parameter space of the SUSY model.
In the NMSSM, the one-loop contributions to the anomalous magnetic moment of the muon or $a_\mu$ are mainly mediated by $\tilde{\chi}^{-}-\tilde{\nu}_\mu$ and $\tilde{\mu}-\tilde{\chi}^{0}$ same as in the MSSM.

The one-loop superpartner contributions to $a_\mu$ can be written as \cite{Martin:2001st, Domingo:2008bb},
\begin{align}
\delta a_\mu^{\tilde{\chi}^0} =& \frac{M_\mu}{16\pi^2} \sum_{\ell=1}^{5} \sum_{m=1}^{2}\Bigg[-\frac{M_\mu}{12M^2_{\tilde{\mu}_m}}\big(\lvert n_{\ell m}^{L}\rvert^2 + \lvert n_{\ell m}^{R}\rvert^2 \big) F_{1}^{N}\big(x_{\ell m}\big) + \frac{M_{\tilde{\chi}_\ell^0}}{3M^2_{\tilde{\mu}_m}}{\rm Re}\big[n_{\ell m}^L n_{\ell m}^R\big]\nonumber\\
&\times F_2^N\big(x_{\ell m}\big)\Bigg]~,\\
\delta a_\mu^{\tilde{\chi}^\pm} =& \frac{M_\mu}{16\pi^2} \sum_{k=1}^{2} \Bigg[\frac{M_\mu}{12M^2_{\tilde{\nu}_\mu}}\big(\lvert c_{k}^{L}\rvert^2 + \lvert c_{k}^{R}\rvert^2 \big) F_{1}^{C}\big(x_{k}\big) + \frac{2 M_{\tilde{\chi}_k^\pm}}{3M^2_{\tilde{\nu}_\mu}}{\rm Re}\big[c_{k}^L c_{k}^R\big] F_2^C\big(x_{k}\big)\Bigg]~,
\end{align}
where the summations label the neutralino and smuon and the chargino mass eigenstates, respectively, and 
\begin{align}
n_{\ell m}^L =& \frac{1}{\sqrt{2}}\big(g_1 N_{\ell 1} + g_2 N_{\ell 2}\big)X_{m1}^{\tilde{\mu}*} - h_\mu N_{\ell 3}X_{m2}^{\tilde{\mu}*}~,\\
n_{\ell m}^R =& \sqrt{2}g_1 N_{\ell 1} X_{m2}^{\tilde{\mu}} + h_\mu N_{\ell 3}X_{m1}^{\tilde{\mu}}~,\\
c_{k}^L =& -g_2 V_{k1}~,\\
c_{k}^{R} =& h_\mu U_{k2}~,
\end{align}
with $h_\mu = \dfrac{M_\mu}{v_d}$ is the muon Yukawa coupling. 
The loop functions are given by 
\begin{align}
F_1^{N}\big(x\big) =& \frac{2}{\big(1-x\big)^4}\Big[1-6x + 3x^2 +2x^3 -6x^2\ln\big(x\big)\Big]~,\\
F_2^{N}\big(x\big) =& \frac{3}{\big(1-x\big)^3}\Big[1-x^2 + 2x\ln\big(x\big)\Big]~,\\
F_1^C\big(x\big) =& \frac{2}{\big(1-x\big)^4}\Big[2+3x-6x^2+x^3 +6x\ln\big(x\big)\Big]~,\\
F_1^C\big(x\big) =& -\frac{3}{2\big(1-x\big)^3}\Big[3-4x+x^2+2\ln\big(x\big)\Big]~, 
\end{align}
where the definition of the variables $x_{\ell m}= M^2_{\tilde{\chi}_\ell^0}/M^2_{\tilde{\mu}_m}$ and $x_{k}= M^2_{\tilde{\chi}_k^\pm}/M^2_{\tilde{\nu}_\mu}$ have been used.
The other diagrams, e.g., mediated by Higgs scalars at one-loop are
  not always numerically significant, though considered in the numerical evaluations
using $\mathtt{NMSSMTools}$.
Similarly, all the dominant two-loop contributions
(Barr-Zee type diagrams)
\cite{Degrassi:1998es, Chang:2000ii, Cheung:2001hz, Chen:2001kn, Arhrib:2001xx, Heinemeyer:2003dq} are included in the $\mathtt{NMSSMTools}$. For instance, the two-loop Higgs diagrams involving a closed SM fermion loop can be important particularly
for a light CP-odd Higgs boson \cite{Domingo:2008bb}.
In the recent past, the anomalous magnetic moment the muon  has been studied in the NMSSM \cite{Cao:2021tuh,Tang:2022pxh,Cao:2022htd}.

\textbf {Anomaly in $W$ boson mass} ($\delta M_W$): The CDF collaboration at the Tevatron \cite{CDF:2022hxs} obtained an updated measurement of the $W$ boson mass, yielding $M_W = 80.4335 \pm 0.0094$ GeV, with a significant improvement in precision compared to the previous measurements and the averages. The recent CDF result, when combined with the former measurements at LEP, the Tevatron, and the LHC, establishes a fresh global average value \cite{deBlas:2022hdk}
    \begin{align*}
    M_W^{\rm Exp} = 80.4133 \pm 0.0080\,\, {\rm GeV}. 
    \end{align*}

    On the contrary, the electroweak precision data within the SM yields \cite{deBlas:2021wap}
    \begin{align*}
        M_W^{\rm SM} = 80.3545\pm 0.0057\,\, {\rm GeV}~,
    \end{align*}
    
    corresponding to a $ 6\sigma$ {\footnote{Note that the central value of the recent measurement by the CDF \cite{Alguero:2022est} is $\sim 7\sigma$ above the SM prediction.}} tension and suggests the possibility of new physics beyond
    the SM. We may note here that the value of $M_W$ has not been settled yet due to incompatibility between the different experimental measurements~\cite{Athron:2022qpo}. \footnote{It may be noted here that the compatibility of $W$ boson
mass measurements,
performed by the ATLAS, LHCb, CDF, and
D0 experiments is further studied considering the theory
uncertainty correlations in Ref. \cite{Amoroso:2023pey}.
The largest theoretical uncertainty in the evaluation of $M_W$ can be observed
from the parton
distribution functions. 
For instance, when combination of all $M_W$
measurements with the CT18 set is considered, one finds
$M_W = 80.3946\pm 0.0115$ GeV. Similarly,
excluding the CDF measurements, one obtains $M_W = 80.3692 \pm 0.0133$ GeV.}. 

The $W$ boson mass predictions in the MSSM \cite{Heinemeyer:2002jq,Heinemeyer:2004gx,Heinemeyer:2006px,Pierce:1996zz,Bagnaschi:2022qhb,Yang:2022gvz,Heinemeyer:2013dia} and its extensions have been detailed in \cite{Domingo:2011uf,Stal:2015zca,Athron:2022isz}. 
The $W$ boson mass, in the presence of the BSM contributions, can be read as,
\begin{align}
M_W^2=\frac{M_Z^2}{2}\left\{1+\left[1-\frac{2\sqrt{2}\pi\alpha}{G_{\mu}M_Z^2}(1+\Delta r)\right]^{1/2}\right\},
\end{align}
where $G_{\mu}$ and $\alpha$ are the Fermi and fine structure constant in the Thomson limit, respectively.
$\Delta r$ refers to all the non-QED radiative contributions to the muon decay, which depends on $M_W$. 
At one-loop $\Delta r$ consists of the $W$ boson self-energy, vertex, box diagrams, and the related counter terms. In the SM, $\Delta r$ can be calculated up to the leading four-loop order, while in the NMSSM, the additional one-loop contributions coming from the Higgs bosons and SUSY particles are evaluated (for analytical expressions, see \cite{Pierce:1996zz, Cao:2008rc}). Similarly, two-loop
diagrams involving Higgs bosons may also be important \cite{Stal:2015zca}. In the numerical analysis, for satisfying the discrepancy $\delta M_W$ between $M_W^{\rm Exp}$ and $M_W^{\rm SM}$ within the NMSSM, we use $\mathtt{NMSSMTools}$. The details of the implementations have been elaborated in Ref. \cite{Domingo:2022pde}. As noted there and also in Ref. \cite{Tang:2022pxh}, with heavier squarks, $\delta M_W$ can receive contributions from comparatively light electroweak sectors, i.e., charginos, neutralinos, left-handed sleptons, and staus.

{\textbf {LHC and LEP bounds on electroweakinos and sleptons :}} Light electroweakinos must satisfy the lower bounds from LEP \cite{OPAL:2002lje, L3:2000euy, ALEPH:2002gap}. In the parameter space where $\tilde{\chi}^\pm_1 \to W^{(*)}+\tilde{\chi}_1^0$ is dominant, constraints from LEP imply $M_{\tilde{\chi}_1^\pm} > 103.5$~GeV. The limit holds for the mass splitting of $\Delta M (\tilde \chi_1^\pm, \tilde \chi_1^0) \geq 3 ~\rm GeV$ while $M_{\tilde{\chi}_1^\pm} > 92.4$~GeV for smaller mass differences. In the recent past, the searches for charginos and sleptons  have been performed at 13~TeV LHC by ATLAS \cite{ATLAS:2018ojr, ATLAS:2018eui, ATLAS:2019lff, ATLAS:2019lng, ATLAS:2020pgy, ATLAS:2019wgx, ATLAS:2021moa} and CMS \cite{CMS:2018kag, CMS:2018szt, CMS:2018eqb, CMS:2020bfa, CMS:2021edw} collaborations.

Usually, in the Higgsino-like LSP models, there are a few combinations of the electroweak states which are of importance in the LHC searches: $\tilde{\chi}^0_1\tilde{\chi}^0_2, \tilde{\chi}^0_1\tilde{\chi}^\pm_1,\tilde{\chi}^0_2\tilde{\chi}^\pm_1,\tilde{\chi}^+_1\tilde{\chi}^-_1$. Following the Ref. \cite{ATLAS:2019lng, ATLAS:2017vat}, Higgsino-like neutralinos or charginos above the LEP limit can be constrained.
For instance, a lower limit can be set at 193 GeV for a mass splitting $\Delta M (\tilde \chi_2^0, \tilde \chi_1^0)$ of 9.3 GeV while a similar limit $M_{\tilde{\chi}^\pm_1}\simeq 150 \rm~ GeV$ for a mass difference of $\sim 3$ GeV \cite{CMS:2021edw} can also be placed from the CMS results. 
The constraints rely on the soft leptons or jets arising in the decays of charginos and neutralinos via off-shell EW gauge bosons $\tilde{\chi}^\pm_1 \to W^{(*)}+\tilde{\chi}^0_1$ and $\tilde{\chi}^0_{2,3,4} \to Z^{(*)}/H_{\rm SM} + \tilde{\chi}^0_1$ ($H_{\rm SM}$ refers to an SM-like Higgs scalar in any BSM model). In the NMSSM, there can be $\tilde{\chi}^0_1 \equiv \tilde B$ or $\tilde S$ dominated, whereas relatively heavier neutralinos $\tilde{\chi}^0_2,\tilde{\chi}^0_3$ may become Higgsino-like with the Higgsino-like lighter
chargino. Even a better-compressed scenario can be conceived when both $\tilde B$ and $\tilde S$ are lighter than the Higgsino-like neutralinos $\tilde{\chi}^0_3,\tilde{\chi}^0_4$.
Now in the presence of a light sleptons/sneutrinos/$H_1$ and the small couplings of the
$\tilde{\chi}^0_1$ \footnote{ In one of our scenarios $\tilde{\chi}^0_1 \equiv \tilde S$ is dominantly singlet-like, thus having a tiny coupling with the SM states.}, the branching fractions for $\tilde{\chi}^\pm_1 \to W^{(*)}+\tilde{\chi}^0_1$ and/or $\tilde{\chi}^0_{2,3,4} \to Z^{(*)}/H_{\rm SM} + \tilde{\chi}^0_1$ can be smaller.
Similarly, lengthy cascade decays into lepton/tau-rich final states may be difficult to detect, which
can lower $M_{\tilde{\chi}^\pm_1}$ in compliance with the LHC results. Since this study is devoted to studying the dependence of the rare $Z$ decay processes on the NMSSM parameter space, we assume, for the parameter space scan, a lighter chargino above $M_{\tilde{\chi}^\pm_1} > 103.5$~GeV, which is not yet completely ruled out. However, the LHC constraints on SUSY searches are verified using $\mathtt{SModelS-2.3.0}$~\footnote{The $\mathtt{CheckMATE}$~\cite{Cacciari:2011ma, Cacciari:2005hq, Cacciari:2008gp, Dercks:2016npn, Read:2002hq}, which uses detailed Monte Carlo simulations, may provide better guidance in studying the validity of any parameter space point of the NMSSM.} \cite{Kraml:2013mwa, Dutta:2018ioj, Khosa:2020zar, Alguero:2021dig}.

We summarise the potentially important final states comprised of $\ell^+\ell^-$ ($\ell \in e,\mu,\tau$) pair, jets, and missing transverse momentum through pair production of charginos, neutralinos, and sleptons, searched at the ATLAS and CMS collaborations. The most recent ones refer to an integrated luminosity of 
139 $fb^{-1}$ at 13 TeV  center-of-mass energy. The recent results, particularly important in the present context, can broadly be classified below \footnote{The new ATLAS and CMS results, relevant in the present context, are included in the recent version of $\mathtt{SModelS-2.3.0}$ \cite{ATLAS:2022hbt, ATLAS:2019lff, ATLAS:2022zwa, ATLAS:2021yqv, CMS:2022vpy, CMS:2022sfi}.}:
\begin{itemize}
    \item 
    $PP\to \tilde{\chi}_1^\pm \tilde{\chi}_2^0 \to Z \tilde{\chi}_1^0 W^\pm \tilde{\chi}_1^0$ ($W$ and $Z$ bosons can be off-shell) \cite{ATLAS:2022zwa,ATLAS:2021moa,ATLAS:2019wgx,ATLAS:2019lng,CMS:2018szt,ATLAS:2018eui}. For on-shell vector bosons, $Z$ and $W$ decay to leptonic and
    leptonic (hadronic) final states, respectively. An ISR jet provides the required handle to detect the soft leptons above the SM background \cite{ATLAS:2019lng, ATLAS:2018eui}.    
  The lower limits for equal-mass $\tilde \chi_1^\pm \tilde \chi_2^0$ are $\sim$ 640-660 GeV for a massless $\tilde \chi_1^0$ while it becomes 300 GeV  for $\Delta M (\tilde \chi_2^0,\tilde \chi_1^0) \simeq M_Z$.
For mass-splitting less than $M_Z$, as mentioned already, $\tilde \chi_1^\pm \tilde \chi_2^0$
decays through off-shell $W, Z$ bosons, the lower limits can go down to 300(100) GeV for $\Delta M (\tilde \chi_2^0,\tilde \chi_1^0) \sim 35(2)~$GeV \cite{ATLAS:2021moa}. The second lightest neutralino may decay to an SM-like Higgs scalar and LSP with 100\% BR \cite{ATLAS:2021moa, ATLAS:2020pgy, CMS:2018szt}. For the $WH_{\rm SM}$ final state, and $\Delta M (\tilde \chi_2^0,\tilde \chi_1^0) \geq M_{H_{\rm SM}}$, one can set a minimum mass, ${M(\tilde \chi_1^\pm, \tilde \chi_2^0)} \sim$ 190 GeV \cite{ATLAS:2021moa}.
    \item
In the parameter space of our concern, all the electroweakinos may well be within the TeV scale. 
Thus the pair production of heavier electroweakinos may be of importance, for instance, when each of them decays into an LSP and an on-shell $W$, $Z$, and SM-like Higgs boson. A mass splitting greater than 400 GeV is considered with scalar and vector bosons further decay hadronically \cite{ATLAS:2021yqv}. Typically, with pair production of dominantly Wino-like states followed by its direct decays into a Bino-like or Higgsino-like LSP, Wino states having masses $ \in (400-1060)$ GeV are excluded when the LSP mass is below 400 GeV, while the mass splitting is $\geq 400$ GeV. On the other hand, in the context of pair production of Higgsino-like states followed by direct decays into a Bino-like or Wino-like LSP, Higgsino-like states with masses $\in (450-900)$ GeV are excluded when the LSP mass is below 240 GeV, and the mass splitting is $\geq 450$ GeV.

\item 
The pair production of lighter charginos $PP\to \tilde \chi_1^\pm \tilde \chi_1^\mp \to W^\pm \tilde \chi_1^0  W^\mp \tilde \chi_1^0 $ have also been considered at the LHC \cite{ATLAS:2019lff,ATLAS:2022hbt}. Further $\chi_1^\mp$ may follow $W$ boson-mediated or slepton-mediated $\tilde \ell \nu (\ell \tilde \nu) \tilde \ell \nu (\ell \tilde \nu) \to \ell \nu  \tilde \chi_1^0 \ell \nu  \tilde \chi_1^0$ decays. For $W$ boson-mediated decays, a comparatively stringent lower bound can be placed that depends on $\Delta M (\tilde \chi_1^\pm,\tilde \chi_1^0)$ \cite{ATLAS:2019lff, ATLAS:2018ojr}.
 \item 
Direct pair-production of sleptons ($\tilde \ell$ refers mainly $\tilde e,\tilde \mu)$ with each decaying into a charged lepton and $\tilde \chi_1^0$, $PP\to \tilde \ell \tilde \ell \to \ell  \tilde \chi_1^0 \ell \tilde \chi_1^0$ have been searched at \cite{ATLAS:2022hbt,ATLAS:2019lff,ATLAS:2019lng,ATLAS:2018ojr,CMS:2018eqb}. A conservative limit on the light-flavored sleptons can be placed using the searches at LEP: $M_{\tilde \mu_R}$ is 94.6 GeV for mass splittings down to $M_{\tilde \mu_R}-M_{\tilde \chi_1^0} \geq 2 ~\rm GeV$ while for $M_{\tilde e_R}$ a universal lower bound $M_{\tilde e_R} \geq 73 ~\rm GeV$ can be placed independent of the mass splitting $\Delta M ({\tilde e_R},{\tilde \chi_1^0})$.  
Typically, for a lighter $\tilde e$ and $\tilde \mu$ with masses $\leq 150$ GeV, the mass splitting $\Delta M ({\tilde \ell}, {\tilde \chi_1^0}) \leq$  
50 GeV is desired for having an acceptable parameter space point. Otherwise, with both $\tilde S$ and/or $\tilde B$ lighter compared to Higgsino-like $\tilde \chi_{3,4}^0$ may reduce the branching fractions in the classical search channels $ \tilde \ell \to \ell  \tilde \chi_1^0 $. With a relatively heavier $\tilde \ell$, one finds $\tilde \ell \to \ell \tilde \chi_i^0$ where $\tilde \chi_i^0$ subsequently decays to LSP through the long cascades. In the case of stau ($\tilde{\tau}$) masses, we require $M_{\tilde{\tau}_1} > 93.2$~GeV following LEP through LHC potentially constraints stau masses from 120 GeV to 390 GeV depending on $\Delta M ({\tilde \tau},{\tilde \chi_1^0})$ \cite{ATLAS:2019gti}.
\end{itemize}

Similarly, the invisible Z decay
     $\Gamma(Z\to \tilde\chi_1^0 \tilde\chi_1^0) < 5 \times 10^{-4} ~\rm GeV$,
    the LEP2 bounds on neutralino production
    $\sigma_Z(e^+e^-\to \tilde\chi_1^0 \tilde\chi_i^0) < 10^{-2}~{\rm pb}~ (i>1)$
    and $\sigma_Z(e^+e^-\to \tilde\chi_i^0 \tilde\chi_j^0) < 10^{-1}~{\rm  pb}~ (i,j>1)$ are always used \cite{DELPHI:2003uqw,OPAL:2003wxm}.
This  may lead to a lower bound on $\lvert\mu_{\rm eff}\rvert \geq 120$ GeV \cite{Das:2010ww}.

\section{Numerical results}
\label{Sec:NR}

Here we present the numerical results for the process $Z\to \phi \gamma$. We perform a scan over the following ranges of the relevant parameters: \\ 
${0.01\le\lambda \le 0.7}$, ${0.001\le\kappa\le 0.2}$, ${6 \le \tan\beta\le 20}$, ${0.1\le A_\lambda \le 2}$, ${-0.2\le A_{\kappa} \le 0}$, 
${0.1\le \mu_{\rm eff} \le 0.3}$.\\
In the above, all the mass parameters are in TeV. For better illustration, we further split
the coupling $\lambda$ into two regimes: (1) 
 the low \textbf{$\lambda$} region: $0.01 \leq \lambda \leq 0.1$~(\textbf{Reg:~1}), and (2) the moderate or high \textbf{$\lambda$} region: $0.1 < \lambda \leq 0.7$ (\textbf{Reg:~2}).
With relatively lower values of $\lambda$ (and $\kappa$), \textbf{Reg:~1} can accommodate a valid Singlino-like DM, while the \textbf{Reg:~2} can have a dominantly Higgsino-like or even a mixed LSP. As discussed, to satisfy direct detection constraints, additional requirements must be fulfilled for a valid DM candidate for the latter parameters. 

\begin{figure}[h]
	\centering
    \includegraphics[width=0.4\linewidth]{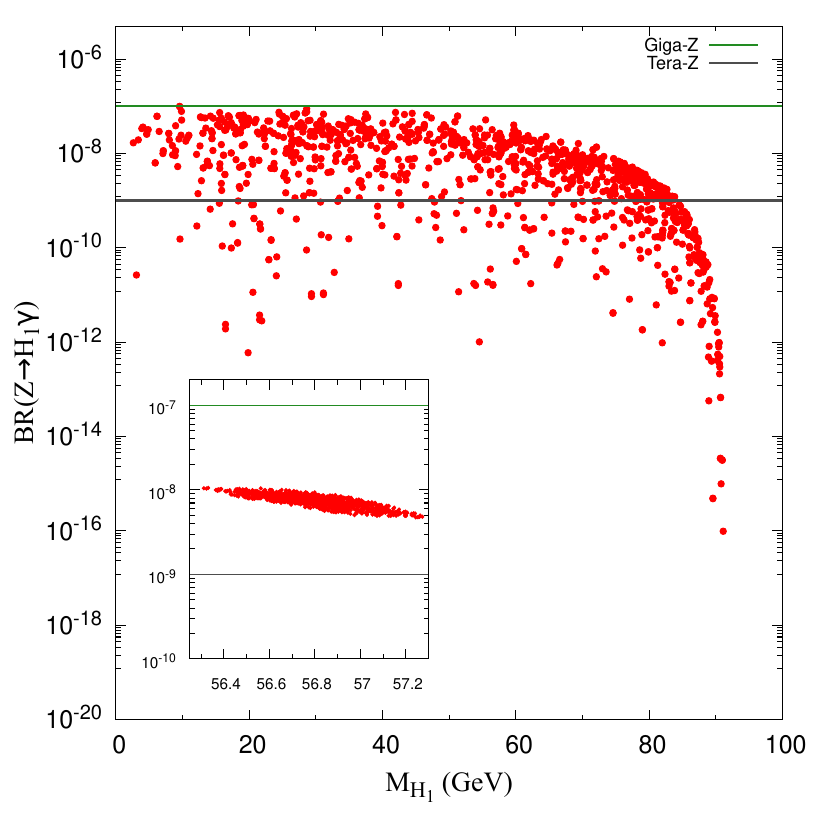}
	\includegraphics[width=0.4\linewidth]{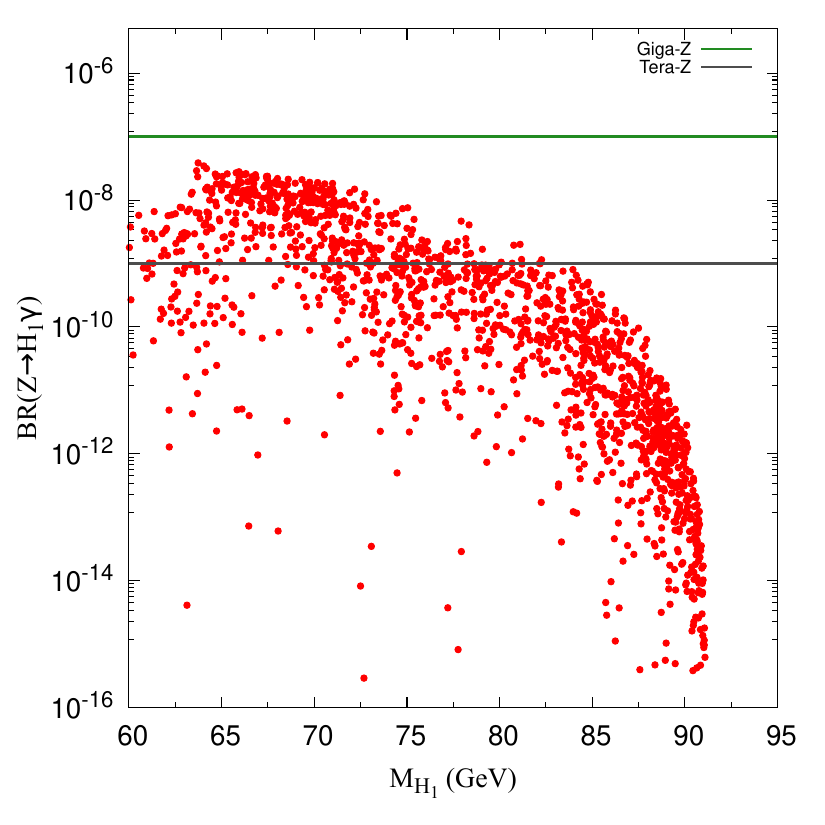}\\
     \hspace{0.85cm}(a)\hspace{5.7cm} (b)\\
    \includegraphics[width=0.4\linewidth]{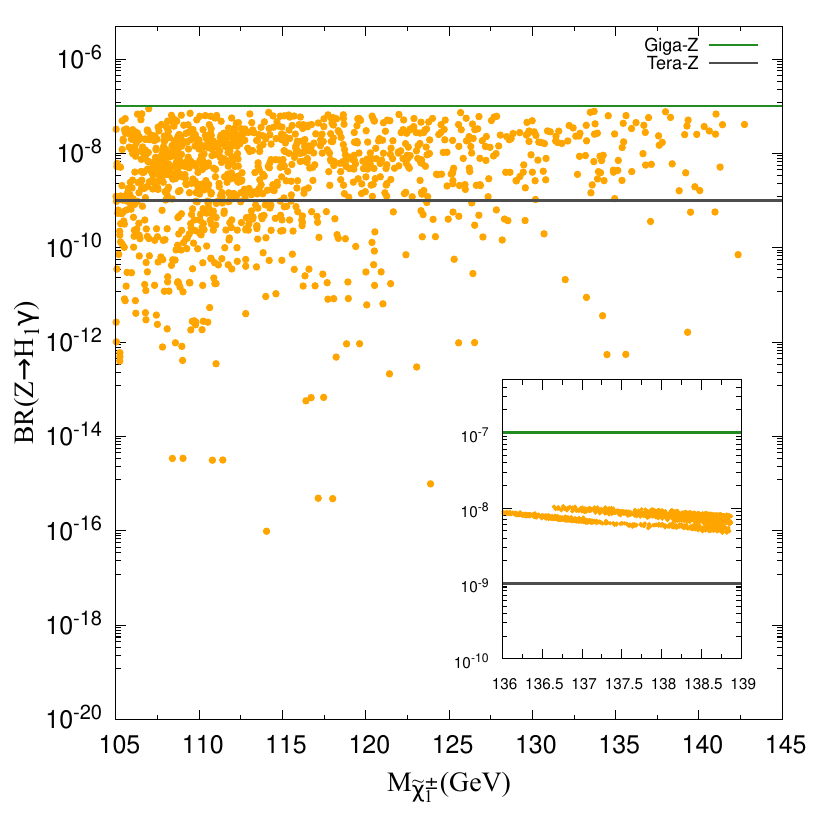}
    \includegraphics[width=0.4\linewidth]{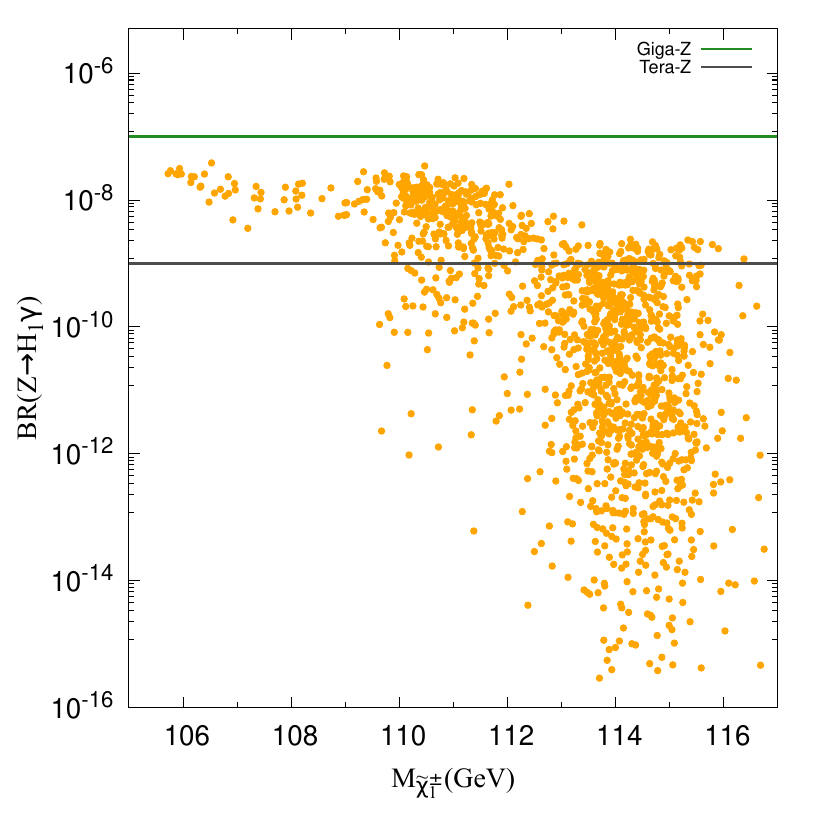}\\
    \hspace{0.85cm}(c)\hspace{5.7cm} (d)
    \caption{Variation of the $BR(Z\to H_1\gamma)$ with $M_{H_1}$ ((a) and (b)) and variation of the $BR(Z\to H_1\gamma)$ with $M_{\tilde{\chi}_1^{\pm}}$ ((c) and (d)) are shown 
    for \textbf {Reg:~1} and 
    \textbf{Reg:~2}, respectively.
    The experimental and theoretical constraints implemented within the $\mathtt{NMSSMTools}$ 
    are respected. The delineated points agree with the two important anomalies associated with $\delta a_\mu$ and $\delta M_W$. All the points in (b) and (d) satisfy $\mathtt{SModelS-2.3.0}$ while the points in the inset
    figures of (a) and (c) obey SUSY searches and DM constraints.}
    \label{br_mcha_mh1}
 \end{figure}

\begin{figure}[h]
	\centering
	\includegraphics[width=0.4\linewidth]{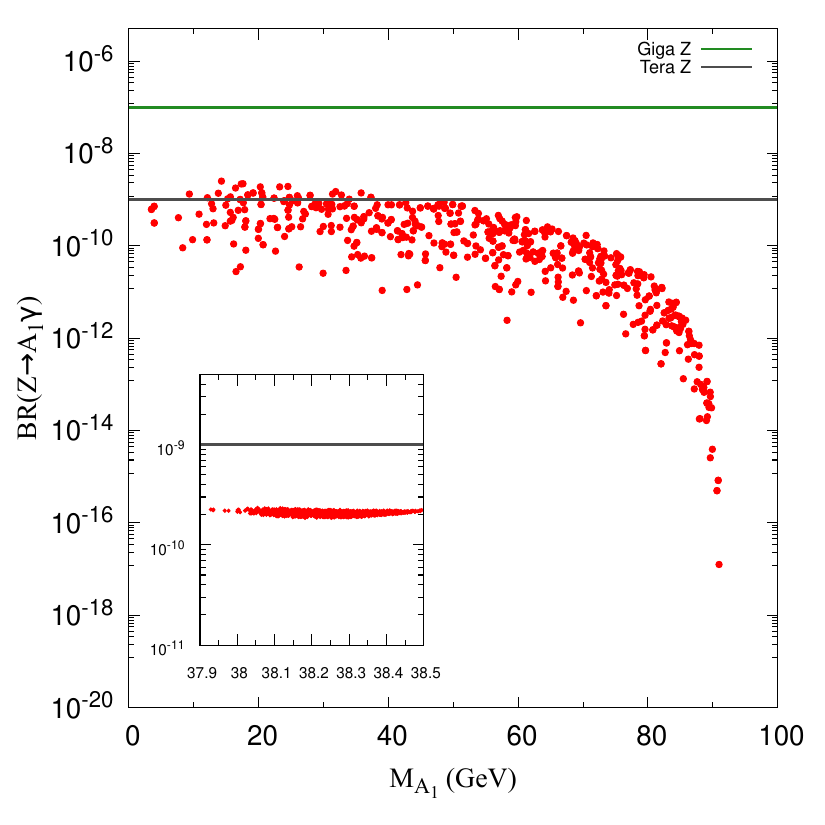}
   \includegraphics[width=0.4\linewidth]{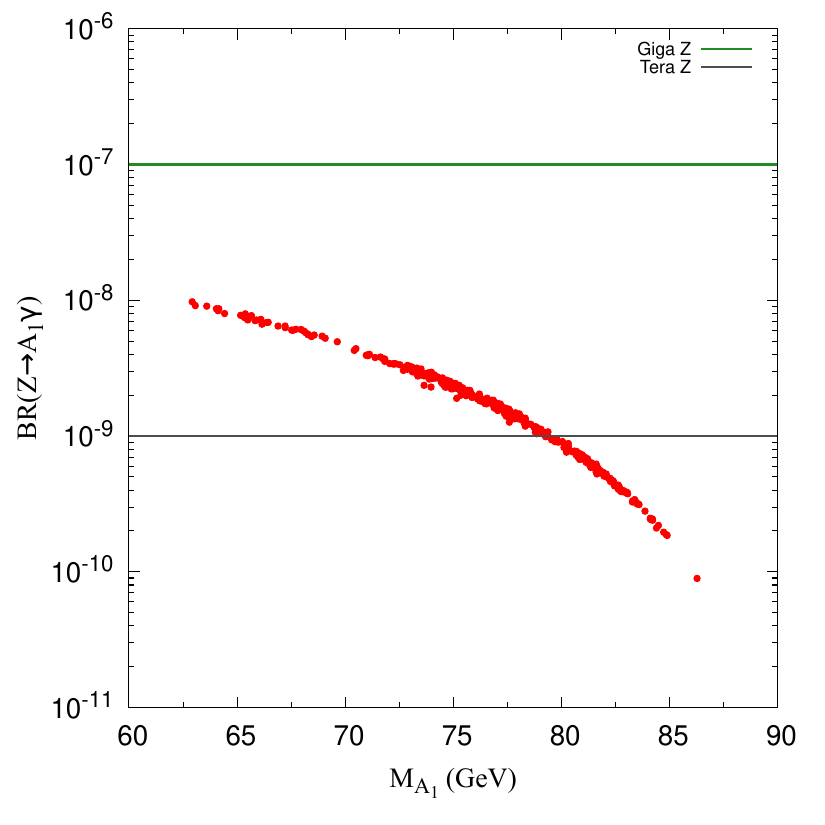}\\
     \hspace{0.85cm}(a)\hspace{5.7cm} (b)\\
    \includegraphics[width=0.4\linewidth]{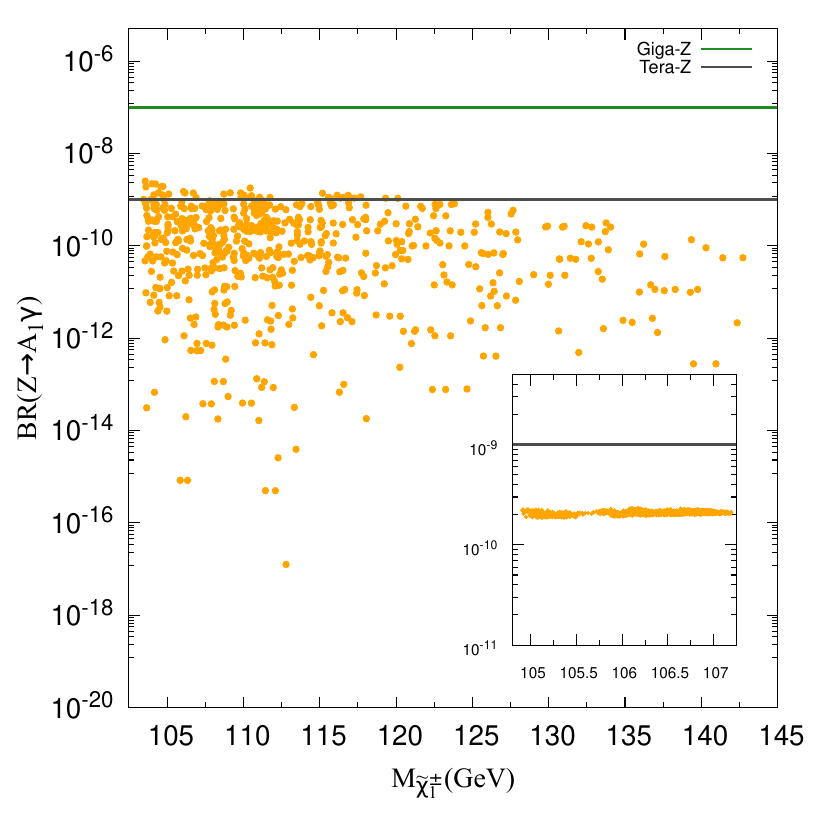}
    \includegraphics[width=0.4\linewidth]{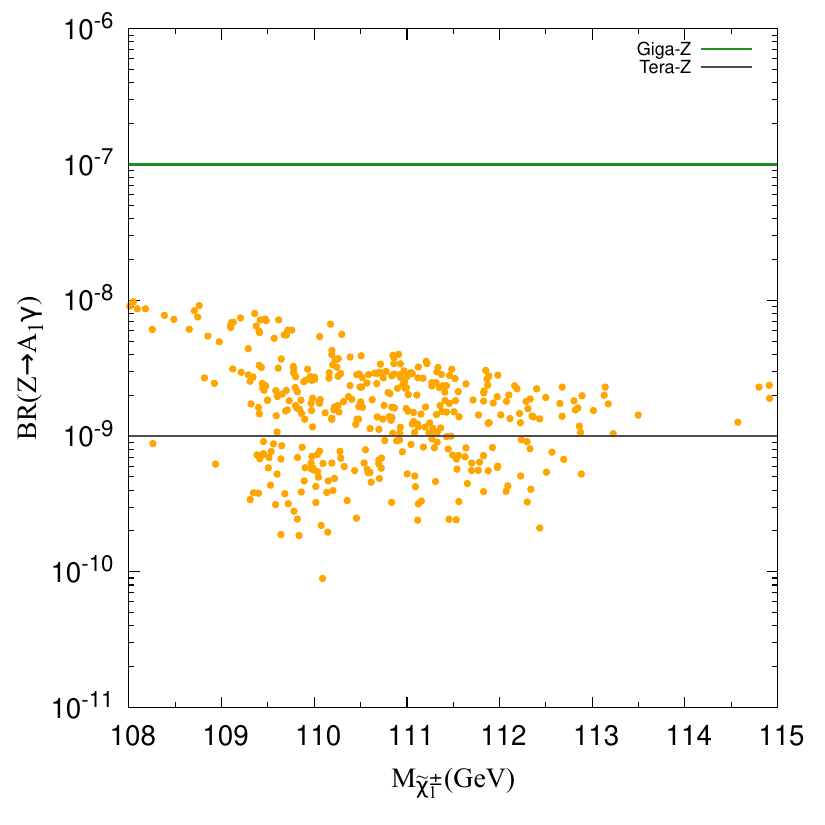}\\
      \hspace{0.85cm}(c)\hspace{5.7cm} (d)\\
      \caption{Same as \ref{br_mcha_mh1}, except the fact that $H_1$ is replaced by the lightest CP-odd scalar $A_1$}
     \label{Fig:zagamma1}
    \end{figure}

We will now try to explore the NMSSM parameter space where a more significant value of the $BR(Z\to \phi \gamma)$ can be observed consistent with the experimental and theoretical constraints implemented within the $\mathtt{NMSSMTools}$.  
The DM constraints for $\tilde \chi_1^0$ are not endorsed as a necessary criterion for a valid parameter space point. The flavor physics constraints and fitting the anomalies to $a_\mu$, $M_W$, through NMSSM contributions
 are always respected.
Additionally, the \textbf{Reg:~2} includes a critical checking of the validity
of each parameter space point under SUSY searches with $\mathtt{SModelS-2.3.0}$. For \textbf{Reg:~1}, fine-tuning among the SUSY parameters will be required to satisfy both DM constraints and limits from SUSY searches. We find it convenient to perform a $\mathtt{MCMC}$ scan within a very small interval in the NMSSM-specific parameters around some fixed values of them. The resultant spectra around a particular input satisfying both DM constraints and SUSY searches are shown in the inset of Fig.~\ref{br_mcha_mh1} ((a), (c)) and   Fig.~\ref{Fig:zagamma1} ((a), (c)).

In Fig.~\ref{br_mcha_mh1}, we present the  branching ratio $BR(Z\to H_1\gamma)$ with the mass of the lightest CP-even scalar ($M_{H_1}$) (top) and the mass of the lightest chargino ($M_{\tilde{\chi}_1^\pm}$) (bottom). 
For instance, in Fig.~\ref{br_mcha_mh1}a, we show the variation of the $BR (Z\to H_1\gamma)$ for relatively lower values of $\lambda$ 
while the same for higher values of $\lambda$ in Fig.~\ref{br_mcha_mh1}b. In \textbf {Reg:~2}, for $M_{H_1}<60$ GeV, $H_2 \to 2 H_1$ may become potentially large and restrictive.  
But in the low $\lambda$ region (\textbf{Reg:~1}), in Fig.~\ref{br_mcha_mh1}a, the $BR(H_2\to 2H_1)$ can be much smaller. Therefore, in this case $M_{H_1}<\frac{M_{H_2}}{2}$ is considered for presentation. In both cases, $ BR(Z\to H_1\gamma)<10^{-9}$ (i.e., below the Tera-$Z$ line) for $M_{H_1}\geq 85$ GeV can be noted, where the phase space suppression coming from Eq.~\eqref{Eq:width_form} is significant. Similarly, a pre-factor boost will help raise the branching fraction when $M_{H_1}$ takes smaller values in \textbf{Reg:~1}. Usually, a smaller value of $\lambda$ reduces the contributions from the charginos in the loop. 
 \noindent
Similarly, the lightest scalar $H_1$ may become
more singlet-like, which can even lower the $W$ boson contributions
in the branching ratio. But a light $\phi$, 
with a phase space boost, can supersede the suppressions driven by the NMSSM couplings; thus,
we may achieve $BR(Z\to H_1\gamma) \leq 10^{-7}$ for smaller values of $M_{H_1} \in (10-40)$~GeV. 
However, when imposing the DM constraints and SUSY search
limits through $\mathtt{SModelS-2.3.0}$, we find a maximum value of the $BR(Z\to H_1\gamma) \leq 10^{-8}$. The region is shown in the inset of Fig.~\ref{br_mcha_mh1}a. The {\textbf{Reg:~2}} also predicts $BR(Z\to H_1\gamma)\leq 10^{-8}$.

In Fig.~\ref{br_mcha_mh1} (bottom), we plot the branching fraction dependence on the mass of the lightest chargino. Compared to \textbf{Reg:~1}, \textbf{Reg:~2} posses relatively 
heavier $H_1$. One finds that most of the parameter space points fall below the Tera-$Z$ line.

The results for the decay $Z\to A_1\gamma$ have been depicted in Fig.~\ref{Fig:zagamma1} for the same ranges of the parameters as above. 
Fig.~\ref{Fig:zagamma1}a and \ref{Fig:zagamma1}b show the variations of the $BR(Z\to A_1\gamma)$ with the mass of the lightest pseudoscalar $M_{A_1}$ in \textbf{Reg:~1} and \textbf{Reg:~2} respectively.
Similarly, Fig.~\ref{Fig:zagamma1}c and \ref{Fig:zagamma1}d show the variations of the $BR(Z\to A_1\gamma)$ with the mass of the lightest chargino $M_{\tilde{\chi}_1^\pm}$. Like the previous case, in Fig.~\ref{Fig:zagamma1}a, $M_{A_1}<60$ GeV is allowed. 
Since the dominant $W$ boson loop is absent, the branching fraction is smaller than that of $BR(Z\to H_1\gamma)$.  
Moreover, the partial width is dominantly controlled by the chargino loops since the SM $t$ quark form factors are usually smaller than the chargino contributions (e.g., see the reference points
in Table \ref{tab:my_label_1}).  

\begin{figure}[h]
	\centering
	\includegraphics[width=0.4\linewidth]{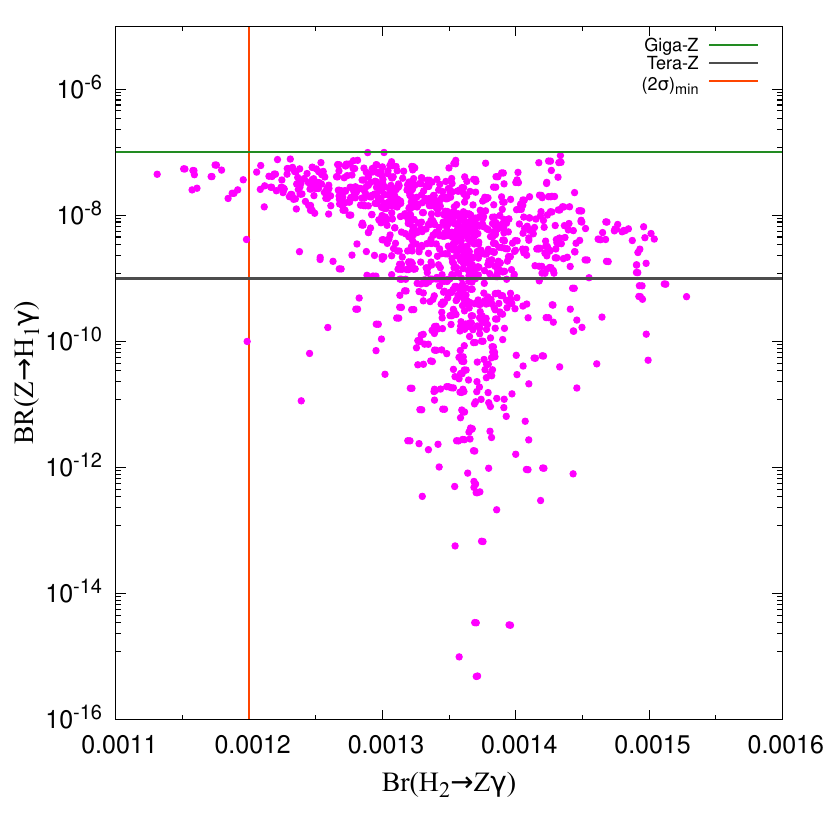}
    \includegraphics[width=0.4\linewidth]{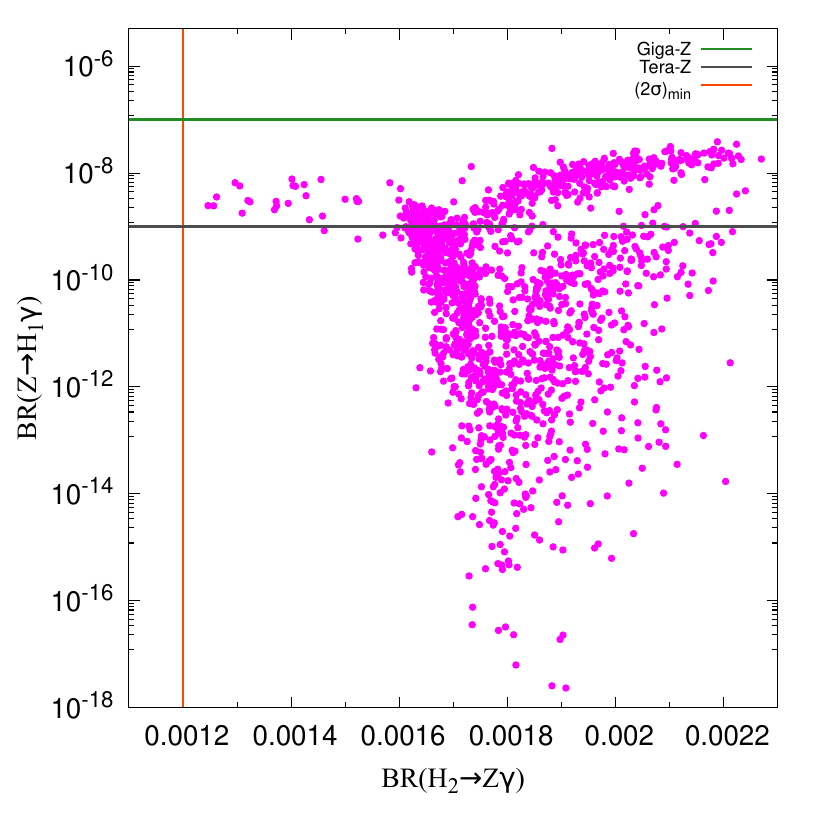}\\
     \hspace{0.85cm}(a)\hspace{5.7cm} (b)\\
    \includegraphics[width=0.4\linewidth]{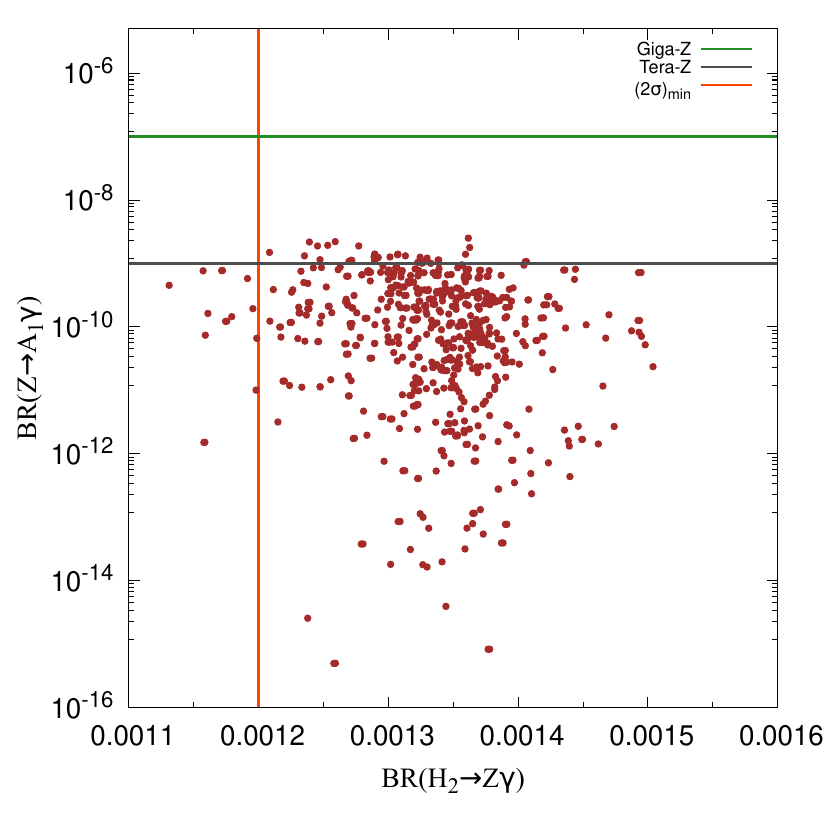}
    \includegraphics[width=0.4\linewidth]{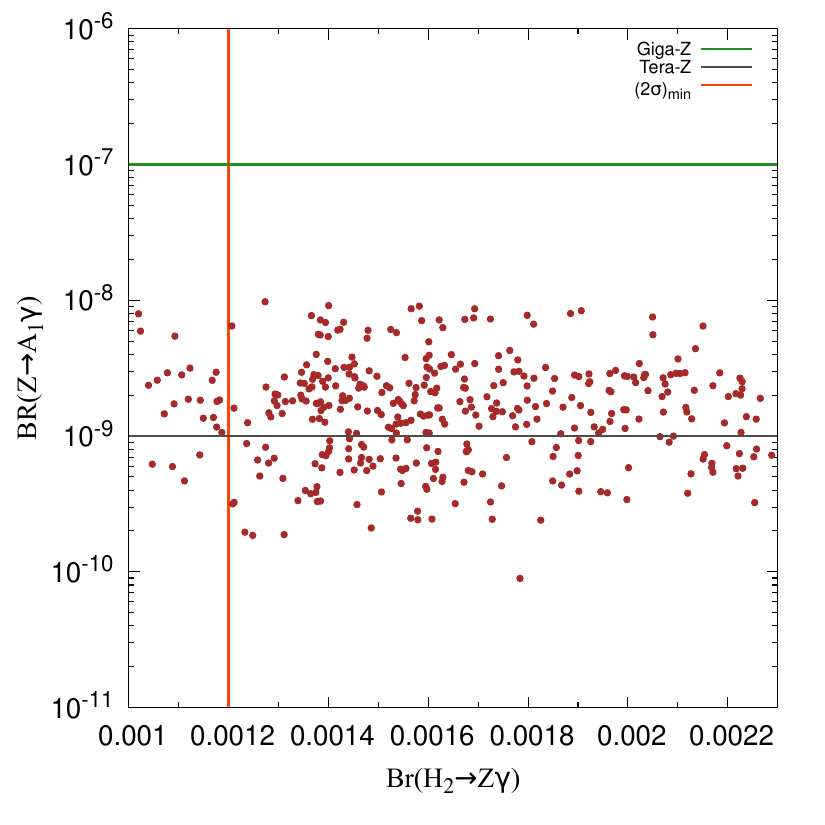}\\
      \hspace{0.85cm}(c)\hspace{5.7cm} (d)
      \caption{Variations of the $BR(Z\to H_1\gamma)$ ((a) and (b)) and $BR(Z\to A_1\gamma)$ ((c) and (d)) with $BR(H_2\to Z\gamma)$ is shown. As in the earlier case, (a) and (c) refer to \textbf{Reg:~1} and (b) and (d) refer to \textbf{Reg:~2}. Here we confine to recent measurements at the LHC, which constrains the branching fraction within $1.2\times 10^{-3} <BR(H_2\to Z\gamma) < 5.6 \times 10^{-3} $. }
\label{Fig:zagamma_H2Zgamma}
\end{figure}
\noindent
So an overall drop in the $Z\to A_1\gamma$ branching fraction can be easily observed. The suppression is more in \textbf{Reg:~1} where small values of $\lambda$ lower the chargino contributions compared to that of \textbf{Reg:~2}. 
This can not be compensated even going to a smaller value for $M_{A_1}$ in Fig.~\ref{Fig:zagamma1}a, where one can avail a significant boost from the kinematic pre-factors. Additionally,
if one applies the DM constraints and SUSY searches, the $BR(Z\to A_1\gamma)$ seems to reduce further (see the inset of
 Fig.~\ref{Fig:zagamma1}a).
Overall, one can achieve a $BR(Z\to A_1\gamma)\sim 10^{-10}$ or
$\sim 10^{-8}$, for \textbf{Reg:~1} and \textbf{Reg:~2} 
respectively.  

Similarly, one can observe from Fig.~\ref{Fig:zagamma1}c and \ref{Fig:zagamma1}d that the $BR(Z\to A_1\gamma)$ decreases gradually with the chargino mass. \textbf{Reg:~2} i.e.,
Fig.~\ref{Fig:zagamma1}d offers better visibility due to a relatively higher $\lambda$ value. This was not clearly observed for the case of $Z\to H_1\gamma$ due to the dominance of the $W$ boson loop.
Since the chargino loops almost solely govern the process $Z\to A_1\gamma$, one can directly measure the NMSSM coupling $\lambda$ by observing the branching fraction.

We now present the parameter space where the recently observed excess on the branching fraction $BR(H_2\to Z\gamma)$ measurements can be co-related with the $BR(Z\to \phi \gamma)$ (see Fig.~\ref{Fig:zagamma_H2Zgamma}). 
We note here that both the observables are dominated by the same loops i.e., $W$ boson and chargino loops. Again we 
choose \textbf{Reg:~1} and \textbf{Reg:~2} to present the variations. The observable $BR(H_2\to Z\gamma)$ is calculated using $\mathtt{NMSSMTools}$. Following the discussion in Sec.~\ref{sec:results} and admitting a 2$\sigma$ variation, the allowance in $BR(H_2\to Z \gamma)$ reads $(1.2-5.6)\times 10^{-3}$.  
Fig.~\ref{Fig:zagamma_H2Zgamma} shows that a significant part of the parameter space, in compliance with $BR(H_2\to Z\gamma)$ allowance, is within the reach of the proposed $e^-e^+$ colliders in the Tera-$Z$ mode.  
Recall that, $BR(Z\to A_1\gamma)$ in low $\lambda$ regions does not offer much hope even in future $Z$ factories. However, the region can be seen to be compatible with $H_2$ decays to $Z$ and $\gamma$, thus, can be probed in the future precision searches of the SM-like Higgs boson. Thus the $BR(H_2\to Z\gamma)$ can offer as an excellent complementary test to the $Z$ boson rare decays.

Having established the necessary parametric dependence of $BR(Z\to \phi \gamma)$ on the NMSSM-specific parameters, we will now have a closer look at the allowed parameter space through a 
few BMPs. Again we follow all the theoretical and experimental checks performed by the $\mathtt{NMSSMTools}$. The SUSY searches are validated by $\mathtt{SModelS-2.3.0}$
while the relevant search channels have already been detailed in Sec.~\ref{sec:results}. 
Guided by the previous lesson, a set of four representative points for studying the prospects of $Z\to H_1 \gamma, A_1 \gamma$ in the NMSSM parameter space for \textbf{Reg:~1} (first and third) and \textbf{Reg:~2} (second and fourth) are shown in Table \ref{tab:my_label}. 
The SUSY input parameters, relevant masses, and mixings for the scalars and fermions are noted therein. The Low-energy observables $\delta a_\mu$ and $M_W$  are also presented. 
Lighter electroweakinos and sleptons are helpful to attune the radiative corrections for having a consistent solution for both of the anomalies.
 Moreover, the sample points can easily fit the recent measurements of the $BR(H_2\to Z\gamma)$ for an SM-like Higgs decay.
 Compared to high $\lambda$ scenarios, in the low $\lambda$ regime, a relatively high value of $\tan\beta$, associated with lighter charginos, can fit the $\delta a_\mu$ anomaly. 
 A relatively high value of $\tan\beta$ boosts the neutralino $\delta a_\mu^{\tilde{\chi}^0}$ and chargino parts $\delta a_\mu^{\tilde{\chi}^\pm}$ in $\delta a_\mu$ calculations. The $\tan\beta$ dependence appears through the muon Yukawa couplings.
As discussed, the LSP in the low $\lambda$ regions
is dominantly Singlino-like with $N_{15}\geq 99\%$, thus satisfying the direct detection limits on the DM. Similarly, we observe a dominantly Higgsino-Singlino-like LSP in \textbf{Reg:~2}. Further, if one lowers $M_2$, the LSP also includes a significant Wino part.
We have also cross-checked the validity of our BMPs with
$\mathtt{HiggsBounds-5.10.2}$.  

\begin{table}[h]
	\centering
     \begin{tabular}{|c|c|c|c|c|c|c|c|c|c|c|c|}
		\hline
		\multicolumn{10}{|c|}{For the lightest CP-even scalar ($H_1$)} \\ 		
		\hline
		BMPs & $\tan\beta$& $\lambda$& $\kappa$ & $A_{\lambda}$& $A_{\kappa}$& $\mu_{\rm eff}$ &$M_{\tilde{\chi}_1^\pm}$  & $M_{\tilde{\chi}_2^\pm}$ & $M_{\tilde{\chi}_1^0}$\\	
		\hline
		  BMP-1 (Small $\lambda$)& 11.00 & 0.0797 & 0.0120 & 2610.00 & $-0.17$ & 188.02 & 132 & 256 & 58\\
        \hline
         BMP-2 (Large $\lambda$)& 6.2 & 0.6090 & 0.2250 & 1005.20 & $-155.97$ & 165.02 & 107 & 238 & 64\\
        \hline
        \hline

		  BMPs& $M_{H_1}$&  $S_{13}$ & $M_{A_1}$&  $P_{13}$ &$M_{\tilde{\ell}_L}$ & $M_{\tilde{\ell}_R}$ &  $\delta a_{\mu}$&$M_W$ & $BR(H_2\to Z\gamma)$  \\	
		\hline
		  BMP-1 (Small $\lambda$)& 56 & 0.994  & 10 & 0.996  & 186 & 204 & $2.04\times 10^{-9}$ & 80.393 & $1.46\times 10^{-3}$ \\
        \hline
         BMP-2 (Large $\lambda$)& 65.5 & 0.994 & 201 & 0.996 & 318 & 318 & $1.63\times 10^{-9}$ & 80.393 & $1.8\times 10^{-3}$ \\
        \hline
        \hline
     \multicolumn{10}{|c|}{For the lightest CP-odd scalar ($A_1$)} \\ 
      \hline

	   BMPs & $\tan\beta$& $\lambda$& $\kappa$ & $A_{\lambda}$& $A_{\kappa}$& $\mu_{\rm eff}$ &  $M_{\tilde{\chi}_1^\pm}$  &$M_{\tilde{\chi}_2^\pm}$ & $M_{\tilde{\chi}_1^0}$ \\	
		\hline
		  BMP-3 (Small $\lambda$)& 15.80 & 0.0367 & 0.0076 & 1370.24 & $-16.204$ & 143.40 &107  & 230 & 61\\
        \hline
         BMP-4 (Large $\lambda$)& 8.53 & 0.5540 & 0.1957 & 1377.24 & 30.008 & 162.80 &108 &234  &63\\
        \hline
        \hline

		  BMPs& $M_{H_1}$&  $S_{13}$ & $M_{A_1}$&  $P_{13}$ & $M_{\tilde{\ell}_L}$ & $M_{\tilde{\ell}_R}$ &$\delta a_{\mu}$ & $M_W$  & $BR(H_2\to Z\gamma)$ \\	
		\hline
		  BMP-3 (Small $\lambda$)& 55 & 0.998  & 38 & 0.999 & 323 & 323  & $3.29\times 10^{-9}$& 80.392 & $1.49\times 10^{-3}$ \\
        \hline
         BMP-4 (Large $\lambda$)& 119 & 0.996 & 64 & 0.998 & 308 & 308 & $2.37\times 10^{-9}$  & 80.408 & $1.57\times 10^{-3}$ \\
        \hline
    
     \end{tabular}
     \caption{Representative points for studying the prospects of $Z\to H_1 \gamma, A_1 \gamma$ in the NMSSM parameter space for the two specific regimes, low and moderate/high values of $\lambda$. The relevant mass terms and mixings of the SUSY particles for our choice of parameters are also listed. Similarly, $\delta a_\mu$ and $M_W$ are shown for the four BMPs. Moreover, the sample points can easily fit the recent measurements of the $BR (H_2\to Z\gamma)$. We have cross-checked the validity of our benchmark points with
       $\mathtt{HiggsBounds-5.10.2}$. Similarly, the LHC constraints on SUSY searches are verified using $\mathtt{SModelS-2.3.0}$ (see text).}
	\label{tab:my_label}
\end{table}

As has been stated, we now numerically manifest the contributions from the different loops of Fig.~\ref{fig:zagamma3} in Table \ref{tab:my_label_1}. The $W$ boson loop with form factor $\mathcal{F}^{(d)}$ provides the leading contributions followed by lighter $\tilde{\chi}_i^\pm$ having form factor $\mathcal{F}^{(a)}$. SM $t$ quark loop ($\mathcal{F}^{(c)}$) comes next in the relative contributions though it interferes destructively with the $W$ boson loop. For CP-odd scalar, $\tilde{\chi}_i^\pm$ completely controls the $Z \to A_1 \gamma$ amplitude through the form factor $\mathcal{\tilde{F}}^{(a)}$. The corresponding branching fraction for $Z\to H_1 \gamma$ or $Z\to A_1 \gamma$ can be noted from Table \ref{tab:my_label_1}. As expected, a relative suppression in $Z\to \phi \gamma$ branching fraction is observed in the low $\lambda$ scenarios.
In general, to accommodate a dominantly Singlino-like LSP to satisfy the DM limits, a dominantly singlet-like $\phi$ ($>99\%$ singlet component) is preferred.  
Although the overall $BR(Z\to \phi \gamma)$ will be diminished, the suppression is more for the $BR(Z\to A_1\gamma)$. We find that the future $Z$-factories can potentially probe our BMPs except the BMP-3.

\begin{table}[h]
	\centering
	\begin{tabular}{|c|c|c|c|c|c|c|c|c|c|}
		\hline
		\multicolumn{8}{|c|}{Form factors arising from the dominant diagrams for CP-even BMPs} \\ 		
		\hline
		     BMPs & $\mathcal{F}^{(a)}$ & $\mathcal{F}^{(b)}$& $\mathcal{F}^{(c)}$ & $\mathcal{F}^{(d)}$& $\mathcal{\tilde{F}}^{(a)}$ & $\mathcal{\tilde{F}}^{(c)}$ & $BR(Z\to H_1\gamma)$\\	
		\hline
		BMP-1 & $2.60\times 10^{-7}$ & $2.40\times 10^{-9}$ & $-3.45\times 10^{-7}$ & $4.70\times 10^{-6}$ & $-$ & $-$ & $1.56\times 10^{-8}$\\
        \hline
        BMP-2 & $3.01\times 10^{-6}$ & $7.70\times 10^{-8}$ & $-4.43\times 10^{-7}$ & $6.77\times 10^{-6}$ & $-$ & $-$ & $3.50\times 10^{-8}$\\
        \hline
        \hline

        \multicolumn{8}{|c|}{Form factors arising from the dominant diagrams for CP-odd BMPs} \\ 		
		\hline
		   BMPs & $\mathcal{F}^{(a)}$ & $\mathcal{F}^{(b)}$& $\mathcal{F}^{(c)}$ & $\mathcal{F}^{(d)}$& $\mathcal{\tilde{F}}^{(a)}$ & $\mathcal{\tilde{F}}^{(c)}$ & $BR(Z\to A_1\gamma)$\\	
		\hline
        BMP-3 & $-$ & $-$& $-$ & $-$ & $3.54\times 10^{-7}$ & $8.31\times 10^{-10}$ & $2.13\times 10^{-10}$\\
        \hline
		BMP-4 & $-$ & $-$ & $-$ & $-$ & $4.69\times 10^{-6}$ & $3.49\times 10^{-8}$ & $1.00\times 10^{-8}$\\
        \hline
	\end{tabular}
	\caption{Contributions from the different loops (see Fig.~\ref{fig:zagamma3}) towards $\mathcal{{F}}$ and $\mathcal{\tilde{F}}$ for the sample points shown in Table \ref{tab:my_label}. In the case of the CP-even Higgs scalar, $\tilde {\chi}_i^\pm$, and $W$ boson provide the leading contributions. SM $t$ quark loop comes next in the relative contributions though it interferes destructively with the $W$ boson loop. For CP-odd scalar, $W$ cannot be present and $\tilde{\chi}_i^\pm$ completely controls the $Z \to A_1 \gamma$ amplitude through the form factor $\mathcal{\tilde{F}}^{(a)}$. }
\label{tab:my_label_1}
\end{table}

We outline here the spectra of electroweakinos and sleptons and their dominant branching fractions into lighter particles. This will, in turn, identify
the characteristics of the spectra for satisfying the important LHC constraints. For instance, the first sample point corresponds to a dominantly singlino-like LSP where $\tilde{\chi}_1^\pm$ decays via off-shell $W$ boson $\tilde{\chi}_1^\pm \to \tilde{\chi}_1^0 + W^*$ with $W^*$ decaying to hadronic or leptonic final states.
The second lightest $\tilde{\chi}_2^0$ decays to $\tilde{\chi}_1^0,H_1$ with $\sim 90\%$ BR. The relatively heavier Higgsino-like and Wino-like mixed neutralinos with masses $\sim (220-250$) GeV decay into sneutrinos and sleptons (including $\tilde \tau_1$). The EW scalars decay through Higgsino-like $\tilde{\chi}_2^0$ with $\tilde{\chi}_2^0$ finally decays to $\tilde{\chi}_1^0$ LSP. The third benchmark point, with a low $\lambda$, shows a similar decay pattern. However, there are a few subtle changes, e.g., $\tilde{\chi}_2^0$ involves three body decays, sleptons decay to EW charginos and heavier neutralinos (mass $\sim 160-280$ GeV) before finally decay to Singlino-like LSP. 
Potentially rich $\tau$ final states may be observed, which can make the LHC searches even more challenging. In the high $\lambda$ scenario, the first sample point refers to the decay pattern of EW particles associated with $Z\to H_1 \gamma$ amplitude. It looks similar to the first sample point, while the heavier chargino and neutralinos appear as the intermediate on-shell states of the sleptons. The last BMP produces $\tau$ rich final states, which may originate through staus in particular. However, with $M_{\tilde \tau_{1,2}} \sim 105-120$ GeV, the sample point is yet to be excluded for our chosen LSP.

\begin{table}[h]
	\centering
	\begin{tabular}{|c|c|c|c|c|c|c|c|c|}
		\hline
		  \multicolumn{2}{|c|}{CP-even (Small $\lambda$)} & \multicolumn{2}{|c|}{CP-even (Large $\lambda$)} & \multicolumn{2}{|c|}{CP-odd (Small $\lambda$)} & \multicolumn{2}{c|}{CP-odd (Large $\lambda$)}\\
      \hline
		 Process & BR & Process & BR & Process &BR &Process &BR\\
		\hline
        \hline
		  $\tilde{\chi}_1^{+}\to \tilde{\chi}_1^0+q+\bar{q}^\prime$& 0.64 & $\tilde{\chi}_1^{+}\to \tilde{\chi}_1^0+q+\bar{q}^\prime$ & 0.66 &$\tilde{\chi}_1^{+}\to \tilde{\chi}_1^0+q+\bar{q}^\prime$  & 0.58 &$\tilde{\chi}_1^{+}\to \tilde{\nu}_{\tau_1}+ \tau^{+}$ & 0.98\\
		\hline
        $\tilde{\chi}_1^{+}\to \tilde{\chi}_1^0+\ell^{+}+\nu_\ell$& 0.36 & $\tilde{\chi}_1^{+}\to \tilde{\chi}_1^0+\ell^{+}+\nu_\ell$ & 0.34 & $\tilde{\chi}_1^{+}\to \tilde{\chi}_1^0+\ell^{+}+\nu_\ell$ & 0.40 & $\tilde{\chi}_2^{0}\to \tilde{\tau}_1^{\mp} + \tau^{\pm}$ & 0.23\\
		\hline
        $\tilde{\chi}_2^0 \to \tilde{\chi}_1^0 + H_1$& 0.93 & $\tilde{\chi}_2^{0}\to \tilde{\chi}_1^0+Z$ & 0.31 & $\tilde{\chi}_2^0 \to \tilde{\chi}_1^0 + \nu_{\tau} + \bar{\nu}_{\tau}$ & 0.61 & $\tilde{\chi}_2^0 \to \tilde{\nu}_{\tau_1} +\bar{\nu}_{\tau}$ & 0.60\\
		\hline
        $\tilde{\chi}_3^{0}\to \tilde{\chi}_1^0+Z$ & 0.34 & $\tilde{\chi}_2^0\to \tilde{\chi}_1^0+H_1$ & 0.65 & $\tilde{\chi}_2^0 \to \tilde{\chi}_1^0 +\tau^{+}+\tau^{-}$ & 0.15 &$\tilde{\chi}_3^0 \to \tilde{\chi}_1^{0} + Z$ & 0.74\\
		\hline
        $\tilde{\chi}_3^{0}\to \tilde{\nu}_{\ell}+\bar{\nu}_\ell$& 0.35 & $\tilde{\chi}_3^0\to \tilde{\chi}_1^0 + Z$ &  0.72  & $\tilde{\chi}_3^0 \to \tilde{\tau}_1^{\mp} + \tau^{\pm}$ & 0.31 &$\tilde{\chi}_4^0 \to \tilde{\chi}_1^{\pm}+ W^{\mp}$ & 0.34\\
		\hline
        $\tilde{\chi}_4^0\to \tilde{\chi}_1^{\pm}+W^{\mp}$& 0.38 & $\tilde{\chi}_4^0 \to \tilde{\chi}_1^{\pm}+W^{\mp}$ & 0.84 & $\tilde{\chi}_3^0 \to \tilde{\nu}_{\tau_1} + \bar{\nu}_{\tau}$ & 0.59 &$\tilde{\chi}_4^0 \to \tilde{\tau}_1^{\mp} + \tau^{\pm}$ & 0.29\\
		\hline
        $\tilde{\chi}_4^0\to \tilde{\ell}^{\pm}+ \ell^{\mp}$&0.37& $\tilde{\ell}_L^{e, \mu} \to \tilde{\chi}_{1,2,4}^0 + e^{-}$ & 0.38 & $\tilde{\chi}_4^0 \to \tilde{\chi}^{\pm} + W^{\mp}$ & 0.47 &$\tilde{\chi}_4^0 \to \tilde{\tau}_2^{\mp} + \tau^{\pm}$ & 0.26\\
		\hline
        $\tilde{\ell}_L^{e, \mu}\to \tilde{\chi}_2^0 + \ell^{e, \mu}$& 0.36 & $\tilde{\ell}_L^{e, \mu} \to \tilde{\chi}_{1,2}^{-} + \tilde{\nu}_\ell^{e,\mu}$ & 0.61 & $\tilde{\chi}_4^0 \to \tilde{\tau}_1^{\mp} +\tau^{\pm}$ & 0.36 & $\tilde{\ell}_L^{e,\mu}\to \tilde{\chi}_{1,2,4}^0 + \ell^{e,\mu}$ & 0.38\\
        \hline
        $\tilde{\ell}_R^{e, \mu}\to \tilde{\chi}_2^0 + \ell^{e, \mu}$& 0.94 & $\tilde{\ell}_R^{e, \mu} \to \tilde{\chi}_{1,4,5}^0 + e^{-}$ &  0.99 & $\tilde{\ell}_L^{e,\mu} \to \tilde{\chi}_{1,2}^{-} + \nu_\ell^{e,\mu}$ & 0.60 & $\tilde{\ell}_L^{e,\mu}\to \tilde{\chi}_{1,2}^{-} + \nu_{\ell}^{e,\mu}$ & 0.61\\
		\hline
        $\tilde{\tau}_{1}^{-}\to \tilde{\chi}_2^0+\tau^{-}$& 0.40  & $\tilde{\tau}_1^{-} \to \tilde{\chi}_{2,4,5}^0 + \tau^{-}$ &  0.52 & $\tilde{\ell}_R^{e,\mu} \to \tilde{\chi}_{2,4,5}^0 + \ell^{e,\mu}$ & 0.98  & $\tilde{\ell}_R^{e,\mu}\to \tilde{\chi}_{1,4,5}^{0} + \ell^{e,\mu}$ & 0.84\\
		\hline
        $\tilde{\tau}_{2}^{-}\to \tilde{\chi}_2^0+\tau^{-}$& 0.84 & $\tilde{\tau}_1^{-} \to \tilde{\chi}_{1,2}^{-} + \nu_{\tau}$ &  0.42 & $\tilde{\tau}_1^{-}\to \tilde{\chi}_2^{0}+ \tau^{-}$ & 0.46 &$\tilde{\tau}_1^{-} \to \tilde{\chi}_1^0 + \tau^{-}$ & 1.00\\
		\hline
        $\tilde{\nu}_\ell^{e, \mu}\to \tilde{\chi}_2^0+ \nu_\ell^{e, \mu}$&0.33  &$\tilde{\tau}_{2}^{-}\to \tilde{\chi}_{4,5}^0+\tau^{-}$   & 0.40  & $\tilde{\tau}_1^{-}\to \tilde{\chi}_1^{-}+ \nu_\tau$ & 0.52 &$\tilde{\tau}_2^{-} \to \tilde{\chi}_1^0 + \tau^{-}$  & 0.68\\
		\hline
        $\tilde{\nu}_\ell^{e, \mu}\to \tilde{\chi}_1^{+} + \ell^{e, \mu}$& 0.66 & $\tilde{\tau}_2^{-} \to \tilde{\chi}_{1,2}^{-} + \nu_{\tau}$ & 0.45 & $\tilde{\tau}_2^{-}\to \tilde{\chi}_{2,3}^0+\tau^{-}$ & 0.53 &$\tilde{\nu}_\ell^{e,\mu} \to \tilde{\chi}_1^{+} + \ell^{e/\mu}$ &0.63\\
		\hline
        $\tilde{\nu}_{\tau_1}\to \tilde{\chi}_{1, 2}^0+\nu_{\tau}$& 1.00 & $\tilde{\nu}_\ell\to \tilde{\chi}_{1,2}^{+} + \ell^{-}$ & 0.62  & $\tilde{\tau}_2^{-}\to \tilde{\chi}_1^{-}+\nu_\tau$ &0.32  &$\tilde{\nu}_{\tau_1} \to \tilde{\chi}_1^0 + \nu_{\tau}$ &1.00\\
		\hline
	\end{tabular}
	\caption{Branching fractions for the relevant electroweak SUSY particles of the BMPs in Table \ref{tab:my_label}.}
	\label{tab:3}
\end{table}

Indeed, the SUSY spectra, with {\it{not-so-well}} separated electroweak particles, the considered $Z$ decay mode can be instrumental in probing the low mass regions of the NMSSM parameter space. Certainly, the proposed high-energy $e^+e^-$ accelerators CEPC, FCC-ee, or ILC are the best bet to search these rare decays of the $Z$ boson. In the next-generation experiments, $Z$ events with several orders of magnitude higher than that of the LEP are likely to be observed \footnote{The LEP limit for $ BR(Z\rightarrow \phi\gamma) \sim 10^{-6}$.}. For instance, we may consider the international $e^+e^-$ Linear Collider (ILC) \cite{Bambade:2019fyw, Erler:2000jg, Baer:2013cma}. In addition to its scheduled running at 250 GeV center of mass energy, it can perform at around $\sqrt{s} = M_Z$, the so-called Giga-$Z$ mode of the ILC. An order of $10^{9}$ or greater $Z$ bosons are expected to be produced, which would improve the expected sensitivity of the rare decays of the $Z$ boson significantly, one of which is $Z\rightarrow \phi\gamma$. Future proposals include circular $e^+e^-$ colliders, CEPC \cite{CEPCStudyGroup:2018rmc, CEPCStudyGroup:2018ghi} and FCC-ee \cite{FCC:2018evy}, plans to run for several years at center-of-mass energy around the $Z$ pole, thus will be operational at the Giga-$Z$ and Tera-$Z$ phases. For the latter, an order of $10^{12}$ of $Z$s will be produced. With recent updates, even more $Z$ boson events are expected to be observed. The branching fraction can be improved from $10^{-6}$ at the LEP to $10^{-8}-10^{-9}$ \cite{Liu:2017zdh, Erler:2000jg},
which are already shown in the Figs.~\ref{br_mcha_mh1}, \ref{Fig:zagamma1} or in \ref{Fig:zagamma_H2Zgamma}. 
As evident, a significant part of the parameter space will be covered in the future $Z$ factories.
Considering $\phi$ to further decay into $\gamma\gamma$, $BR(Z\to\phi\gamma) \sim 3 \times 10^{-9}$ can be probed for $M_{\phi} \le 100$~GeV \cite{Liu:2017zdh}. 
A similar limit can be derived if $\phi$ decays to leptonic final states \cite{CEPCPhysicsStudyGroup:2022uwl}. If $\phi$ is a pseudo-scalar and decays to a pair of b-jets, see Ref. \cite{Cacciapaglia:2021agf}. For our choices of $M_\phi$, the signal-background differentiation may become difficult. For the HL-LHC probes on exotic $Z$ decays, see \cite{Liu:2017zdh}. 

The rare $Z$ decay processes open up another pathway to produce a singlet-like scalar or pseudoscalar at the 14 TeV LHC because of the huge production cross-section for the $Z$ boson at the leading order, $\sigma_Z\simeq 55.6$ nb. Therefore, one can roughly have a production cross-section for the singlet-like scalar or pseudoscalar, $\sigma_{\phi}=\sigma_{Z}\times BR(Z\to \phi\gamma)\simeq 10^{-8}\times 55.6$ nb $\simeq 0.556$ fb for $M_\phi\leq M_Z$. Further with $\phi$ decays to $b\bar{b}$ (90\% BR, as in BMP-2 and BMP-4), one finds $\Big[\sigma(PP\to \phi)\times {\rm BR}\Big]_{\rm NMSSM}\simeq 0.5$ fb. Certainly, this can be interesting to directly produce $\phi$ at the high-luminosity LHC ($\sqrt{s}=14$ TeV, $\mathcal{L}=3\,\, {\rm ab}^{-1}$).

\section{Conclusions}
\label{Sec:conclusion}
We have studied the rare decay processes $Z\to H_1\gamma$, $A_1\gamma$ in the context of the Next-to-Minimal Supersymmetric Standard Model. Starting with a detailed analytical depiction of the processes through a model-independent parametrization of the new physics couplings, we examine the NMSSM parameter space where the branching fraction $Z\to \phi \gamma$ is important. Analytical expressions from the loop diagrams are implemented in $\mathtt{NMSSMTools}$ to facilitate our analysis. Then we have considered two regimes corresponding to low and high values of $\lambda$, referred to as \textbf{Reg:~1} and \textbf{Reg:~2} respectively. 
The lower values of $\lambda$ (and $\kappa$) offer a dominantly Singlino-like dark matter with a dominantly singlet-like $\phi$ (singlet part is more than 99\%) to satisfy the DD bounds.
The chargino loops become much suppressed due to the smallness of $\lambda$. Similarly, the SM-loop contributions appeared to be small. So the $Z\to \phi \gamma$ amplitude resides on the lower side 
compared to the high $\lambda$ scenarios in \textbf{Reg:~2} where the DM constraints are relaxed on the available parameter space. 
For the latter case, we may observe a larger branching ratio for the rare $Z$ decays which we are interested in.
Apart from satisfying the theoretical and experimental constraints, including SUSY searches, we aim to fit $(g-2)_\mu$, $W$ boson mass anomaly. The spectra are endowed with the recent measurements of the $BR(h_{\rm SM}\to Z\gamma)$, which is $\sim$ 2.2 times larger than the SM prediction, based on the recent measurement of ATLAS and CMS.
Future proposals such as ILC, CEPC, and FCC-ee are anticipated to operate for multiple years, focusing on center-of-mass energy near the $Z$ pole. Consequently, these projects will be capable of conducting experiments at the Giga-$Z$ ($10^{9}$ of $Z$ bosons) and Tera-$Z$ ($10^{12}$ of $Z$ bosons) phases, which can probe a significant part of the NMSSM parameter space. In the BMPs, we find that
	except for BMP-3, all other representative points can be probed in the Tera-$Z$ run. 
	However, BMP-3, a Singlino-like LSP scenario accompanied by compressed electroweak spectra, the $Z\to A_1\gamma$ observable, supplemented by $BR(H_2 \to Z \gamma)$ can also be tested in the near future. 
	Overall, $BR(h_{\rm SM}\to Z\gamma)$ may act as an important
        standby test to $Z\to H_1\gamma$, $A_1\gamma$ in the NMSSM parameter space,
        if it
        gets settled to any deviation compared to that of the SM in the future
        runs of the LHC. Finally, we want to emphasize here that a large
        branching fraction for $Z\to H_1\gamma$, $A_1\gamma$ along with the
        SUSY contributions to $(g-2)_\mu$, $W$ boson mass anomaly can
        be boosted in the presence of the lighter charginos. Additionally,
        lighter
        Higgs bosons are needed, particularly for the $BR(Z\to \phi\gamma)$. So, in the parameter space with low $\mu_{\rm eff}$ and with relatively smaller values of the
        NMSSM-specific trilinear coupling parameters, observance of the
        $BR(Z\to \phi\gamma)$ can serve as a nice complementary test to
        probe the NMSSM-specific signatures compatible
        with the EW precision measurements.
       Additionally, considering the huge production
        cross-section for the $Z$ boson
	at the LHC, this rare decay process can be used
	for the production of singlet-like scalars or pseudoscalars at the hadron colliders.

\section{Acknowledgements}
Our computations were supported in part by SAMKHYA: the High-Performance Computing Facility provided by the Institute of Physics (IoP), Bhubaneswar, India. We acknowledge Anirban Kundu, Amit Chakraborty, and Pradipta Ghosh for their valuable discussions.

\section{Appendix}
\label{sec:appendix}
\subsection{SM and NMSSM vertices}
\label{sec:appendixA}
{\textbf{\underline{SM vertices:}}}
\begin{align*}
&Z_\mu t\bar{t} : \frac{g_2}{4c_W} \gamma_\mu \Big[\Big(1-\frac{8}{3}s_W^2\Big)-\gamma_5\Big]\\
&A_\mu f \bar{f} : eQ_f\gamma_\mu\\
&Z_\mu(p_1)W_\alpha^{+}(p_2)W_\beta^{-}(p_3) : -g_2c_W \big[\big(p_1-p_2\big)_\beta\, g_{\mu\alpha} + \big(p_2-p_3\big)_\mu\, g_{\alpha\beta} + \big(p_3-p_1\big)_\alpha\, g_{\beta\mu}\big]\\
&A_\mu(p_1)W_\alpha^{+}(p_2)W_\beta^{-}(p_3) : -e \big[\big(p_1-p_2\big)_\beta\, g_{\mu\alpha} + \big(p_2-p_3\big)_\mu\, g_{\alpha\beta} + \big(p_3-p_1\big)_\alpha\, g_{\beta\mu}\big]\\
\end{align*}
{\textbf{\underline{NMSSM vertices:}}}
\begin{align*}
&H_aH^\pm H^\mp : \frac{\lambda^2}{\sqrt{2}}\Bigl\{s \big(\Pi_a^{311}+\Pi_a^{322}\big)-v_d \Pi_a^{212}-v_u \Pi_a^{112}\Bigr\} + \sqrt{2}\lambda\kappa s \Pi_a^{312}+ \frac{\lambda A_\lambda}{\sqrt{2}}\Pi_a^{312}\\
&\hspace{2.0cm}+\frac{g_1^2}{4\sqrt{2}}\Bigl\{v_u\big(\Pi_a^{211}-\Pi_a^{222}\big) + v_d \big(\Pi_a^{122}-\Pi_a^{111}\big)\Bigr\} + \frac{g_2^2}{4\sqrt{2}}\Bigl\{v_d\big(\Pi_a^{111}+\Pi_a^{122}\\
&\hspace{2.0cm}+2\Pi_a^{212}\big)+ v_u\big(\Pi_a^{211}+\Pi_a^{222} +2\Pi_a^{112}\big)\Bigr\} + \frac{\lambda \mu^\prime}{\sqrt{2}}\Pi_a^{312}\\
&H_i W_\mu^{\pm}W_\nu^{\mp} : g_{\mu\nu} \frac{g_2^2}{\sqrt{2}}\big(v_d S_{i1}+v_u S_{i2}\big)\\
&H_i(p)H^{\pm}(p^\prime)W_\mu^\mp : \frac{g_2}{2} \big(\cos\beta S_{i2}-\sin\beta S_{i1}\big)\big(p-p^\prime\big)_\mu\\
&A_i(p)H^{\pm}(p^\prime)W_\mu^\mp : i\frac{g_2}{2} \big(\cos\beta P_{i2}+\sin\beta P_{i1}\big)\big(p-p^\prime\big)_\mu\\
&Z_\mu H^{\pm}(p)H^{\mp}(p^\prime) : \frac{g_1^2-g_2^2}{\sqrt{g_1^2+g_2^2}}\big(p-p^\prime\big)_\mu\\
&H_at_Lt_R^c : -\frac{M_t}{\sqrt{2}v\sin\beta}S_{a2}\\
&A_at_Lt_R^c :  -i\frac{M_t}{\sqrt{2}v\sin\beta}P_{a2}\\
&H_a\tilde{\chi}_i^{\pm}\tilde{\chi}_j^{\mp} : \frac{\lambda}{\sqrt{2}}S_{a3}U_{i2}V_{j2} + \frac{g_2}{\sqrt{2}}\big(S_{a2}U_{i1}V_{j2}+S_{a1}U_{i2}V_{j1}\big)\\
&A_a\tilde{\chi}_i^{\pm}\tilde{\chi}_j^{\mp} : i\Bigg[\frac{\lambda}{\sqrt{2}}P_{a3}U_{i2}V_{j2} - \frac{g_2}{\sqrt{2}}\big(P_{a2}U_{i1}V_{j2}+P_{a1}U_{i2}V_{j1}\big)\Bigg]\\
&Z\chi_\ell^{\pm}\chi_m^{\mp} : \frac{g_2}{c_W}\gamma_\mu\big(\mathcal{O}^L_{m\ell}\mathbb{P}_L + \mathcal{O}^R_{m\ell}\mathbb{P}_R \big), {\rm where}\,\,\,\, \mathcal{O}^L_{m\ell}= -V_{m1} V^{*}_{\ell 1}-\frac{1}{2} V_{m2} V^{*}_{\ell 2} + \delta_{m\ell}s_W^2,\,\, \\&\hspace{2cm}\mathcal{O}_{m\ell}^R=-U^{*}_{m1} U_{\ell 1}-\frac{1}{2} U^{*}_{m2} U_{\ell 2}+\delta_{m\ell}s_W^2,\,\, \mathbb{P}_L=\frac{1-\gamma_5}{2}\,\,{\rm and}\,\, \mathbb{P}_R=\frac{1+\gamma_5}{2}\\
&A_\mu\tilde{f}_r(p)\tilde{f}_s(p^\prime) : e(p-p^\prime)_\mu\delta_{rs}\\
&Z_\mu\tilde{f}_r(p)\tilde{f}_s(p^\prime) : \frac{g_2}{2c_W}(p-p^\prime)_\mu\Bigg[\sum_{i=1}^{3}W_{ir}^{\tilde{f}*}W_{is}^{\tilde{f}}-2s_W^2\delta_{rs}\Bigg]\\
&H_1\tilde{f}_{La}\tilde{f}_{La} : \sqrt{2}\Bigg[h^2_{ea}v_uS_{11}+\Big(-\frac{g_1^2}{4}+\frac{g_2^2}{4}\Big)(v_uS_{12}-v_dS_{11})\Bigg]\\
&H_1\tilde{f}_{Ra}\tilde{f}_{Ra} : \sqrt{2}\Bigg[h^2_{ea}v_dS_{11}+\frac{g_1^2}{2}(v_uS_{12}-v_dS_{11})\Bigg]\\
&H_1\tilde{f}_{La}\tilde{f}_{Ra} : \frac{h_{ea}}{\sqrt{2}}(A_{ea}S_{11}-\mu_{\rm eff}S_{12}-\lambda v_uS_{13})\\
&H_1\tilde{\tau}_1\tilde{\tau}_1 : \cos^2\theta_L g_{H_1\tilde{f}_{L3}\tilde{f}_{L3}} + \sin^2\theta_L g_{H_1\tilde{f}_{R3}\tilde{f}_{R3}} + 2\cos\theta_L\sin\theta_L g_{H_1\tilde{f}_{L3}\tilde{f}_{R3}}\\
&H_1\tilde{\tau}_2\tilde{\tau}_2 : \sin^2\theta_L g_{H_1\tilde{f}_{L3}\tilde{f}_{L3}} + \cos^2\theta_L g_{H_1\tilde{f}_{R3}\tilde{f}_{R3}} - 2\cos\theta_L\sin\theta_L g_{H_1\tilde{f}_{L3}\tilde{f}_{R3}}\\
&H_1\tilde{\tau}_1\tilde{\tau}_2 : \cos\theta_L\sin\theta_L \big(g_{H_1\tilde{f}_{R3}\tilde{f}_{R3}} -g_{H_1\tilde{f}_{L3}\tilde{f}_{L3}}\big) + \big(\cos^2\theta_L-\sin^2\theta_L\big)g_{H_1\tilde{f}_{L3}\tilde{f}_{R3}}\\
&A_\mu H^\pm(p) H^\mp(p^\prime): -e\big(p-p^\prime\big)_\mu
\end{align*}

\subsection{Loop amplitudes using Feynman parameters}
\label{Eq:loop_feynman}
Fig.~\ref{fig:zagamma3}a can be calculated in terms of the Feynman parameters as 
\begin{align}
\mathcal{F}^{(a)} = &\frac{i}{16\pi^2}eg_{Z\tilde{\chi}_\ell^\pm\tilde{\chi}_m^\mp}g_{H_1\tilde{\chi}_\ell^\pm\tilde{\chi}_n^\mp}\delta_{mn}K^{(+)}_{LR} \int_{0}^{1}dx\int_{0}^{1-x}dy\Bigg[\frac{N_{\tilde{\chi}^{\pm}}(x,y)}{\Delta_{\tilde{\chi}^{\pm}}(x,y)}\Bigg]~,\\
\mathcal{\tilde{F}}^{(a)} = &-\frac{1}{16\pi^2}eg_{Z\tilde{\chi}_\ell^\pm\tilde{\chi}_m^\mp}g_{A_1\tilde{\chi}_\ell^\pm\tilde{\chi}_n^\mp}\delta_{mn}K^{(+)}_{LR}\int_{0}^{1}dx\int_{0}^{1-x}dy\Bigg[\frac{\tilde{N}_{\tilde{\chi}^{\pm}}(x,y)}{\Delta_{\tilde{\chi}^{\pm}}(x,y)}\Bigg]~,
\label{Fig:1afeyn}
\end{align}
where $N_{\tilde{\chi}^{\pm}}(x,y) = 4\Big[M_\ell-\big(1-x-y\big)\big(M_m+2M_\ell+M_n\big)+\big(M_m-M_\ell\big)\big(1-x\big)+2\big(1-x\big)\big(1-x-y\big)\big(M_\ell+M_n\big)\Big]$, $\tilde{N}_{\tilde{\chi}^{\pm}}(x,y) = 4\Big[-M_\ell+ \big(M_m-M_n\big)\big(1-x-y\big)+\big(M_\ell-M_m\big)\big(1-x\big)\Big]$, and $\Delta_{\tilde{\chi}^{\pm}}(x,y) = x\big(x-1\big)M_Z^2 + xM_\ell^2 +yM_m^2+\big(1-x-y\big)M_n^2 + x\big(1-x-y\big)\big(M_Z^2-M_\phi^2\big)$.\\

For the $t$ quark loop, we can write
\begin{align}
   \mathcal{F}^{(c)}
= &\frac{ig_2e}{64\pi^2c_W }N_c Q L_{LR}^{(+)}g_{H_1tt}M_t \int_{0}^{1}dx\int_{0}^{1-x}dy\Bigg[\frac{N_{t}(x,y)}{\Delta_{t}(x,y)}\Bigg]~,\\
\tilde{\mathcal{F}}^{(c)} = &\frac{g_2e}{64\pi^2c_W }N_c Q L_{LR}^{(+)}g_{A_1tt}M_t\int_{0}^{1}dx\int_{0}^{1-x}dy\Bigg[\frac{\tilde{N}_{t}(x,y)}{\Delta_{t}(x,y)}\Bigg]~,
\end{align}
where $N_{t}(x,y) = 4\Big[1-4x(1-x-y)\Big]$, $\tilde{N}_{t}(x,y) = 1$, and $\Delta_{t}(x,y) = x(x-1)M_Z^2+M_t^2+ x\big(1-x-y\big)\big(M_Z^2-M_\phi^2\big)$.
Similarly, Fig.~\ref{fig:zagamma3}d can be expressed in terms of Feynman parameters as 
\begin{align}
  \mathcal{F}^{(d)} = &-\frac{ieg_2 c_W}{16\pi^2} g_{H_1 W^\pm W^\mp}  \int_{0}^{1}dx\int_{0}^{1-x}dy\Bigg[\frac{N_{W^\pm}(x,y)}{\Delta_{W^\pm}(x,y)}\Bigg]~,
\end{align}
where $N_{W^{\pm}}(x,y) = 2\Big[10x\big(x+y-1\big)-y+5\Big]$ and $\Delta_{W^\pm} = x\big(x-1\big)M_Z^2+M_W^2 + x\big(1-x-y\big)\big(M_Z^2-M_\phi^2\big)$.

The generic result for the diagrams mediated by the scalars is given by
\begin{align}
    \mathcal{F}^{(b,e,f)} = -\frac{i}{16\pi^2}eg_{ZS_1S_2}g_{H_1S_1S_3}\delta_{S_2S_3}\int_{0}^{1}dx\int_{0}^{1-x}dy \Bigg[\frac{N_{S}(x,y)}{\Delta_{S}(x,y)}\Bigg]~,
\end{align}
where $N_{S}(x,y) = 8x(1-x-y)$ and $\Delta_{S}(x,y) = x(x-1)M_Z^2 + x M_{S_1}^2+y M_{S_2}^2 + (1-x-y)M_{S_3}^2+ x\big(1-x-y\big)\big(M_Z^2-M_\phi^2\big)$.

\bibliographystyle{JHEPCust.bst}
\bibliography{zdecay}
\end{document}